\documentclass[letterpaper,twocolumn,10pt]{article}
\usepackage{usenix,epsfig,endnotes}

\usepackage{color,colortbl}
\usepackage{hyperref}
\usepackage{amsmath,amsopn,amssymb,subfloat,xurl}
\usepackage{listings}
\usepackage{subcaption}
\usepackage{endnotes,microtype,xspace,graphicx,fancyvrb,multirow}
\usepackage{booktabs}
\usepackage{array,underscore,relsize}
\usepackage[T1]{fontenc}
\usepackage{times}
\usepackage{fancyhdr,lastpage}
\usepackage{enumitem}
\usepackage[labelfont=bf,font=small,skip=5pt]{caption}
\usepackage{fp}
\usepackage{siunitx}

\usepackage{balance}

\usepackage{menukeys}
\usepackage{titlesec}
\usepackage{pdfpages}

\usepackage{lscape}

\newcommand{\cc}[1]{\mbox{\smaller[0.5]\texttt{#1}}}

\input{code/fmt}

\def\Snospace~{\S{}}

\makeatletter
\renewcommand{\itemautorefname}{\@gobble}
\makeatother

\usepackage{pifont}
\definecolor{darkgreen}{rgb}{0, 0.5, 0}
\newcommand{\yes}{\textbf{\textcolor{darkgreen}{\ding{52}}}\xspace}
\newcommand{\no}{\textbf{\textcolor{red}{\ding{55}}}\xspace}

\newif\ifdraft\drafttrue
\newif\ifnotes\notestrue
\ifdraft\else\notesfalse\fi

\input{glyphtounicode}
\newcolumntype{R}[1]{>{\raggedleft\let\newline\\\arraybackslash\hspace{0pt}}p{#1}}

\newcommand{\squishlist}{
\begin{itemize}[noitemsep,nolistsep]
  \setlength{\itemsep}{-0pt}
}
\newcommand{\squishend}{
  \end{itemize}
}

\usepackage{tikz}

\usepackage{xstring}
\newcommand{\PP}[1]{
\vspace{2px}
\noindent{\bf \IfEndWith{#1}{.}{#1}{#1.}}
}

\newcommand{\boxbeg}{
\vspace{2px}
\noindent\begin{tabular}{|l|}\hline
\begin{minipage}{3.2in}
\vspace{2px}
\noindent
}

\newcommand{\boxend}{
\vspace{2px}
\end{minipage}\\ \hline
\end{tabular}
\vspace{-10pt}
}

\definecolor{codegreen}{rgb}{0,0.6,0}
\definecolor{codegray}{rgb}{0.5,0.5,0.5}
\definecolor{codepurple}{HTML}{C42043}
\definecolor{backcolour}{HTML}{F2F2F2}
\definecolor{bookColor}{cmyk}{0,0,0,0.90}

\usepackage{textcomp}
\lstdefinestyle{mystyle}{
    language=SQL,
    backgroundcolor=\color{backcolour},
    commentstyle=\color{codegreen},
    keywordstyle=\color{codepurple},
    numberstyle=\numberstyle,
    stringstyle=\color{codepurple},
    basicstyle=\scriptsize\ttfamily,
    breakatwhitespace=false,
    breaklines=false,
    captionpos=b,
    keepspaces=true,
    numbers=left,
    numbersep=-0pt,
    showspaces=false,
    showstringspaces=false,
    showtabs=false,
    morekeywords={boolean, BOOLEAN, REFERENCES, WITHOUT, ROWID, IS, TEXT, RENAME, TO, SHOW, INDEXES, RECOVER, JOB, FUNCTION, RETURNS, SERIAL, READS, SQL, DATA,
    RETURN, PARAM_MARKER, IF, PARTITION, HASH, FOR, UPGRADE},
    aboveskip=0pt,
    belowskip=0pt
}
\lstdefinestyle{onlycommenthl}{
    language=SQL,
    backgroundcolor=\color{backcolour},
    commentstyle=\color{codegreen},
    keywordstyle=\color{codepurple},
    numberstyle=\numberstyle,
    stringstyle=\color{codepurple},
    basicstyle=\scriptsize\ttfamily,
    breakatwhitespace=false,
    breaklines=false,
    captionpos=b,
    keepspaces=true,
    numbers=left,
    numbersep=-0pt,
    showspaces=false,
    showstringspaces=false,
    showtabs=false,
    morekeywords={boolean, BOOLEAN, REFERENCES, WITHOUT, ROWID, IS, TEXT, RENAME, TO, SHOW, INDEXES, RECOVER, JOB, FUNCTION, RETURNS, SERIAL, READS, SQL, DATA,
    RETURN, PARAM_MARKER},
    aboveskip=0pt,
    belowskip=0pt,
    keywordstyle=\scriptsize\ttfamily,
    stringstyle=\scriptsize\ttfamily,
    identifierstyle=\scriptsize\ttfamily
}
\newcommand\numberstyle[1]{%
    \scriptsize
    \color{codegray}%
    \ttfamily
    \ifnum#1<10 0\fi#1 %
}

\newcommand{\etal}{\textit{et al}.\xspace}
\newcommand{\ie}{\textit{i}.\textit{e}.\xspace}

\newcommand\aflpp{\cc{AFL++}\xspace}
\newcommand\libfuzzer{\cc{LibFuzzer}\xspace}

\newcommand\sqlsmith{\cc{SQLsmith}\xspace}
\newcommand\sqlancer{\cc{SQLancer}\xspace}
\newcommand\squirrel{\cc{Squirrel}\xspace}
\newcommand\sqlright{\cc{SQLRight}\xspace}
\newcommand\dynsql{\cc{DynSQL}\xspace}
\newcommand\QPG{\cc{QPG}\xspace}
\newcommand\sqlancerqpg{\cc{SQLancer}\textsubscript{\cc{+QPG}}\xspace}

\newcommand\sqlite{\cc{SQLite}\xspace}
\newcommand\mysql{\cc{MySQL}\xspace}
\newcommand\postgres{\cc{PostgreSQL}\xspace}
\newcommand\mariadb{\cc{MariaDB}\xspace}
\newcommand\cockroachdb{\cc{CockroachDB}\xspace}
\newcommand\tidb{\cc{TiDB}\xspace}

\newcommand\norec{\cc{NoREC}\xspace}
\newcommand\pqs{\cc{PQS}\xspace}
\newcommand\tlp{\cc{TLP}\xspace}

\newcommand{\sys}{\mbox{\cc{ParserFuzz}}\xspace}
\newcommand{\sysnocov}{\mbox{\cc{ParserFuzz}\textsubscript{\cc{-cov}}}\xspace}

\newcommand{\cockroachdbsqlsmith}{\mbox{\cc{SQLsmith}\textsubscript{\cc{C}}}\xspace}
\newcommand{\gosqlsmith}{\mbox{\cc{SQLsmith}\textsubscript{\cc{G}}}\xspace}

\newcommand{\totalbugs}{81\xspace}
\newcommand{\crashes}{29\xspace}
\newcommand{\assertbugs}{52\xspace}
\newcommand{\fixed}{34\xspace}

\newcommand{\STAB}[1]{\begin{tabular}{@{}c@{}}#1\end{tabular}}

\begin{document}

\date{}

\title{\Large \bf Parser Knows Best: \\ Testing DBMS with Coverage-Guided Grammar-Rule Traversal}

\author{
    \rm
    Yu Liang $^\dagger$\; \ \
    Hong Hu $^\dagger$\; \ \ 
    \\\\
    \textit{$^\dagger$Pennsylvania State University}
}

\maketitle

\thispagestyle{empty}

\begin{abstract}

\noindent
Database Management System (DBMS) 
is the key component for data-intensive applications. 
Recently,
researchers propose many tools
to comprehensively test DBMS systems
for finding various bugs. 
However,
these tools only cover a small subset
of diverse syntax elements
defined in DBMS-specific SQL dialects,
leaving a large number of
features unexplored.

In this paper,
we propose \sys,
a novel fuzzing framework
that automatically extracts grammar rules
from DBMSs' built-in syntax definition files
for SQL query generation.
Without any input corpus,
\sys can generate diverse query statements
to saturate the grammar features of the tested DBMSs,
which grammar features could be missed by previous tools. 
Additionally,
\sys utilizes code coverage as feedback
to guide the query mutation,
which combines different DBMS features
extracted from the syntax rules
to find more function and safety bugs.
In our evaluation,
\sys outperforms all state-of-the-art existing DBMS testing tools
in terms of bug finding,
grammar rule coverage and code coverage. 
\sys detects \totalbugs previously-unknown bugs
in total across 5 popular DBMSs,
where all bugs are confirmed and \fixed have been fixed.

\end{abstract}

\section{Introduction}
\label{s:intro}

Database Management System (DBMS)  
stores, retrieves and manages
data in a structured manner. 
They are extensively used
in real-world data-intensive applications
to drive trillions of Internet services
and electronic devices%
~\cite{sqlite-famous,mysql-users,
  postgres-users,cockroachdb-users,
  tidb-users,
  stonebraker2013intel}.
Any DBMS bugs will affect a large number of
users~\cite{magellan2,restaurant-breach, 
caesar-breach}.

Recent efforts on DBMS testing%
~\cite{zhong:squirrel,sqlsmith,ba2023testing,
jiang2023dynsql}
can be classified into two categories:
generation-based testing and
mutation-based grey-box fuzzing.
\sqlsmith~\cite{sqlsmith}
is the most popular
generation-based testing tool
to date.
It generates SQL queries
based on pre-defined query templates. 
These templates are manually crafted
by \sqlsmith's  developers,
and can help generate high-quality 
SQL queries~\cite{sqlsmith-bugs}.
Another representative
generation-based tool is \sqlancerqpg,
where \QPG represents \textit{query plan guidance}~\cite{ba2023testing}. 
\sqlancerqpg adopts DBMS query plan information
as the feedback to guide the query generation process,
and is designed to detect logic errors from
the DBMS code.
Similar to \sqlsmith,
\sqlancerqpg also generates query sequences
based on pre-defined SQL templates. 
However,
due to the significant difference
between multiple DBMS dialects,
these pre-defined SQL templates
cannot cover unique, complicated features
from different DBMS systems%
~\cite{liang2022sqlright, msrags}. 
Further,
as every DBMS keeps evolving,
developers of \sqlsmith and \sqlancerqpg
have to track all recent updates
and manually insert new templates.
%

Recently,
coverage-guided grey-box fuzzing is widely used
to detect memory errors from a wide-range of 
applications, including but not limited to 
operating systems~\cite{chointfuzz,syzkaller,moonshine,kim2020hfl,xu:janus},
web browsers~\cite{clusterfuzz,xu:freedom,shou2021corbfuzz,guo2013gramfuzz}
and compilers~\cite{chen:polyglot,dinh:favocado,park:die}.
To conduct grey-box testing,
a fuzzing tool, or \textit{fuzzer},
first instruments
the target program to record code coverage. 
Then,
it generates a large number of random inputs
and executes the target program.
If one input triggers new code coverage,
the fuzzer will treat the input as interesting
and save it to a queue. 
By accumulating interesting inputs
from the fuzzing queue,
the fuzzer aims to trigger as many code logic
from the target program as possible.
Eventually,
some inputs will trigger memory errors
in the target program.

Researchers have adopted grey-box fuzzing
to test DBMS systems~\cite{zhong:squirrel,
wen2023squill,jiang2023dynsql,chen:polyglot}.
For example,
\squirrel performs syntax-preserving mutation
and semantics-guided instantiation
to generate high-quality SQL queries%
~\cite{zhong:squirrel}.
It adopts code coverage
to guide the input mutation and scheduling.
However,
the diversity of \squirrel-generated queries
heavily relies on the quality of provided seed corpus%
~\cite{liang2022sqlright,rebert2014optimizing,moonshine}.
The more features the seed corpus covers,
the more diversity the mutated inputs could trigger,
and therefore,
the more bugs can be discovered.
If the input corpus misses one grammar feature,
any bug related to this feature
would be missed by \squirrel. 
However,
collecting a feature-rich seed corpus
is a challenging task.
The unit test suites
provided from the target program
may only cover parts of all features.
Although \squirrel can generate queries
with high validity rate,
it can hardly explore the vast SQL features
provided by various DBMS dialects.
Therefore,
how to effectively
explore the vast syntax features
in various SQL dialects
is the key challenge to achieve
efficient fuzzing.

A parser generator reads the grammar rules from a syntax definition file, and then leverage 
the grammar rules to a source code that can analysis any inputs according to the defined 
syntax rules~\cite{johnson1990yacc, parr2011ll, parr1995antlr}.
If parsing successfully, the parser source code would transform the input into a parser 
tree, where the tree nodes represent the parsed tokens coming from the raw input. 
The developer is free to further transform the parser trees into their software internal 
structures.
The main purpose of the parser is to verify the syntax correctness of
the provided input, and then leverage the raw input into the program's
internal structure for further operations.  
In our study, most of the popular DBMSs today use parser generators
to define their SQL grammar rules~\cite{mysql-parser, mariadb-parser, 
cockroachdb-parser, tidb-parser, 
sqlite-parser, postgres-parser},
including all the DBMSs we tested in this paper.
Intuitively, these grammar rules provided for the parser generators would 
cover all the possible syntax regulations for the target programs, because the parsers
work as verifiers to check all input sources in the software front-ends. 
Therefore, the grammar definition files for the parser generators can serve as 
an uniform interface that we can use to explore the ground truth SQL grammar
rules defined for the dedicated DBMSs.

In this paper, we introduce a new query generation and mutation method, 
that directly learn and apply the grammar rules from the DBMSs' built-in parsers
to the fuzzing process.
Instead of using the grammar rules from DBMS parsers to simply verify the input queries,
we use the parser rules to generate and mutate random query statements for fuzzing. 
By traversing all the grammar rules defined for the DBMS built-in parsers,
the generated queries could saturate all the interesting 
syntax features supported by the original DBMS programs. 
Furthermore, we combine the code coverage feedback from the traditional grey-box fuzzing
to guide the query grammar-based mutation.
If the fuzzed query triggers a new code coverage from the DBMS program,
we save the parsed syntax tree to the fuzzing queue
for further query mutations.

We implemented a new fuzzing tool, \sys, that automatically constructs 
interesting query sequences based on DBMS built-in parser rules, and 
then uses the code coverage feedback to guide the query syntax rule-based
mutation. 
We tested \sys on 5 most popular open-source DBMSs,
\mysql, \sqlite, \cockroachdb, \tidb and \mariadb. 
\sys found \totalbugs bugs from all 5 DBMSs, which contain \crashes 
segmentation faults and \assertbugs assertion failures. 
\sys achieves the largest detected bug numbers, the highest grammar coverage and 
code coverage 
compared to the other state-of-the-art DBMS testing tools in our evaluations.

In summary, this paper makes the following contributions.

\squishlist

\item We propose a novel method
  for automatic SQL-query generation and mutation,
  which utilizes existing syntax rules of DBMS parsers
  to randomly generate feature-rich
  statements, without relying on high-quality seed corpus. 

\item We utilize code coverage as the feedback
  to select the promising grammar rules
  for generating new queries.
  The coverage-guided rule-based generation
  will help explore deep code logic
  of complicated DBMS programs. 

\item We evaluate \sys on 5 real-world DBMSs
  and find \totalbugs new bugs.
  We demonstrate that our tool covers more grammar features,
  which helps trigger more memory errors
  than state-of-the-art DBMS testing tools.

\squishend


\PP{Open Source}
We will release the source code of \sys
at \url{https://github.com/SteveLeungYL/parserfuzz_code}
to help protect popular DBMS systems.

\section{Background \& Challenges}
\label{s:background}


\subsection{An Example Memory Error from MySQL}
\label{ss:bg-example}

\autoref{l:example} shows a query that triggers a segmentation fault on the release
version of MySQL DBMS (version 8.0.33). 
This bug was detected by our newly proposed tool, \sys, without any seed corpus that
explore on the similar syntax. 
The proof of concept (PoC) is constructed with two simple lines of SQL queries. 
The first query creates a new temporary table named \cc{t0} with a column called \cc{c1}.
Temporary tables are only available within the current server-client connected session, 
and the data in the temporary table would be automatically dropped when the
client exits the connection.
In this PoC, the \cc{CREATE TABLE} statement appends an index to 
the table to enhance the speed of data retrieval. 
Interestingly, the created index is a \cc{composite index}, 
which means the created index is constituted with two different parts of data. 
The first part contains one single column.
The second part is a functional key parts index, often referred as functional index.
Different from normal index, where only columns are considered as keys.
Functional index uses query expression as the indexed
content, and the indexed expression receives a speed up.
Functional index is useful when the DBMS user constantly employs the same
SQL expression to search for table data. 
After caching the expression into index,
the DBMS can speed up the repeated expression handling in 
the data retrieval. 
In the PoC of \autoref{l:example}, the index \cc{i2} contains
a mixed usage of normal index and functional index. 
In the second statement, the PoC promptly looks for the 
created index information.

When running this PoC on the latest version of \mysql (version 8.0.33), the
targeted DBMS encounters \cc{NULL pointer dereference} crash in the temporary 
table handling logic. 
The problem arises from the mixed usage of normal index and functional
index, which confuses the temporary table creation handling, and
the expression \cc{c1 + c1} is not correctly saved in the index 
creation.
When the \cc{SHOW INDEX} query searches for the stored index information, 
it fails to find the columns saved into the functional index (being NULL), 
and therefore results in a crash. 
The PoC can be easily weaponized to upload to any online
\mysql services and conduct \cc{Denial-of-Service} attack. 
We have reported the bug to the \mysql developers.
The developer has confirmed the bug and marked the bug severity as \cc{Serious}.

\begin{figure}[t]
\begin{lstlisting}[style=mystyle,label={l:example},caption={\textbf{A segmentation fault crashing from \mysql} due to the mixed usage of \cc{normal index} and \cc{functional index}. }]
CREATE TEMPORARY TABLE t0(c1 INT, INDEX i2(c1, (c1+c1)));
SHOW INDEXES IN t0;
\end{lstlisting}
\end{figure}

The PoC shown in \autoref{l:example} is interesting because,
although previous works like \squirrel\cite{zhong:squirrel} have been extensively tested on \mysql,
none of the previous tools found this bug. 
The key to uncovering this bug is to combine normal index creation and functional
index creation in a single \cc{CREATE TEMPORARY TABLE} statement. 
However, the internal representation of \squirrel doesn't supports using expressions
as index keys, nor does \squirrel includes functional index 
in its input seeds. 
Therefore, not matter how hard \squirrel mutates on its input, it will never detect
this bug. 
Furthermore, there are only one single instance in the \mysql official unit test that 
explores the combined use of normal index and functional index. 
%
And the unit test is not constructed with temporary table. 
As such, the DBMS tester is easy to overlook this one line unit test example and thus not able to 
include any queries that combine these two index types. 
Therefore, all the previous testing tools missed this bug.
\sys automatically learns all the syntax features from the ground truth \mysql syntax 
definition file. 
Without relying on any pre-existing unit test corpus, it can craft queries that
explore the rarely tested feature such as combined use of normal index and functional index.
As a result, our tool \sys finds this bug in 10 hours of \mysql fuzzing.

\subsection{Generation-based DBMS Testing Tools}

There are some existing DBMS testing tools that relies on SQL 
templates to generate the SQL queries for testing. 
The most well-known generation-based query testing framework is
\sqlsmith~\cite{sqlsmith}.
Since its release, \sqlsmith has been used extensively
to test on different DBMSs, and found many bugs from the DBMS
softwares~\cite{sqlsmith-bugs}. 
However, the hand-written query templates are limited in 
covering the syntax elements from the DBMS syntax rules, and cannot fit in 
the complex and ever-changing SQL dialects defined in different
DBMS softwares.
Therefore, the queries generated from \sqlsmith cannot cover
all the SQL features from the DBMSs, and lack the capability to detect
deep and unique bugs that are dedicated to the DBMSs' feature sets.

There is another generation-based DBMS testing tool called \sqlancer\cite{sqlancer}, that
focuses on detecting DBMS logic errors from DBMS systems~\cite{rigger:pqs, rigger:tlp, rigger:norec}. 
DBMS logic bugs are code logic errors that cause the DBMS to return
incorrect results. 
Unlike tools that detect memory errors, \sqlancer doesn't 
generate arbitrary types of random queries for fuzzing. 
Instead, it focuses only on generating queries that 
matching its oracles' needs. 
In essence, \sqlancer prefers to generate multiple
syntactically different, but functionally equivalence queries, 
and verify their results to ensure the query execution correctness.
\sqlancer introduces a few SQL oracles for this purpose such as 
\norec, \tlp and \pqs, where each shares a distinct SQL pattern 
to match~\cite{rigger:pqs, rigger:tlp, rigger:norec}.
With its latest configuration \sqlancerqpg~\cite{ba2023testing}, it uses the DBMS query plan to guide its
query generation in order to stress test the DBMS query optimization logic.
However, because \sqlancer and \sqlancerqpg
restricted themselves to generate queries
that align with the oracles' patterns, they lack the query diversity needed 
to explore all the grammar features provided by the DBMSs. 
Therefore, neither \sqlancer nor \sqlancerqpg are suitable to detect DBMS memory corruption bugs 
that are arise from interesting but rarely tested syntax features.

There are several other generation-based DBMS testing tools that
aim to detect various kinds of bugs from the DBMSs~\cite{mishra:gen,j:modelfuzz,QAGen}.
Some existing works treat the SQL query generation as a boolean satisfiability problem,
and use SAT solvers to produce queries that achieve high correctness rate~\cite{solver-lj,solver}.
Chandra \etal extend the database construction technique
to boost the efficiency of DBMS query testings~\cite{extra1}.
Bruno \etal propose to generate queries based on Cardinality Constraints~\cite{bruno2006generating}.
On the topic of performance issues, researchers also propose several tools to detect performance bugs 
in the DBMS~\cite{extra2,bovenzi2012aging,yang2018not}.
\cc{APOLLO} runs the same query on different versions of the same DBMS to detect performance regression
bugs~\cite{jung:apollo}.
\cc{AMOEBA} generates functional equivalence queries and checks whether they finish in a similar
response times~\cite{liu2022automatic}. 
Lastly, regarding logic errors, differential testing is employed to detect logic bugs in DBMSs
~\cite{msrags}.
These testing tools check the result consistency from identical queries when running them on different 
DBMSs~\cite{sparkfuzz} or running them in one DBMS but with different
versions~\cite{snowtrail}.
If any result discrepancy is observed, a potential logic bug is found.

\subsection{Mutation-based DBMS Testing Tools}

\squirrel is one of the state-of-the-art mutation-based query testing tools.
It supports testing 4 DBMSs including \sqlite, \postgres, \mysql
and \mariadb~\cite{zhong:squirrel}. 
The core idea of \squirrel is two-folded. 
The first contribution to transform any inputted
queries into its internal representation (IR). 
Based on the generated IR, \squirrel performs type-based mutations 
on the IR tree nodes, in order to mutate the parsed queries
but preserve the mutated queries' syntactic correctness.  
The second contribution is to build a dependency graph 
for the query arguments such as table names and column names.
By resolving the dependency graph, \squirrel fills in the query arguments 
and assigns the parsed query with special semantic meanings.
However, even though \squirrel uses an internal parser to convert the
raw query input into its IR, it still heavily depends on the
provided seed corpus to shape the mutated queries. 
If the given seeds lack a specific SQL grammar, 
\squirrel will not generate any inputs to explore this
grammar feature. 
Given the complexity of the DBMS SQL dialects, it is challenging for \squirrel
to cover all the interesting features from its seed corpus. 
Therefore, the DBMS features that \squirrel explores are limited.
Additionally, \dynsql~\cite{jiang2023dynsql} implements another DBMS fuzzing tool.
It samples the DBMS execution states after every query execution, aiming to 
gather more real-time feedback from the DBMS to improve the
generated queries' correctness rate. 
The sampled query state later guides the query generation,
helping to sidestep some semantic errors caused by the previous unsuccessful
data creations or modifications. 
The idea of \dynsql is complementary to our work.
While \dynsql focuses on generating queries with higher correctness rate,
\sys strives to generate more diverse queries that can explore 
more rarely tested features.
There are a few other mutation-based fuzzing tools for detecting DBMS crashing 
bugs~\cite{wang2013model,li2022fuzzing}.
For example, \cc{RATEL} targets enterprise-level DBMSs 
such as \cc{GaussDB}~\cite{wang2021industry}.
\cc{LEGO} instantiates the query statements with \cc{type-affinity} awareness 
for higher query correctness rate~\cite{liang2023sequence}.
Unfortunately, \dynsql, \cc{RATEL} or \cc{LEGO} have not released their source code, 
which means it is not feasible to compare our work to theirs.


\subsection{Parser Generators in the DBMSs}

A parser generator program takes a grammar definition file as input and
generates source code that parses any input characters according to
the rules defined in the grammar file~\cite{johnson1990yacc, parr2011ll, parr1995antlr}.
Most parser generators use a syntax notion type similar to the \cc{Backus–Naur form} 
(\cc{BNF})~\cite{BNFforms, bisonformat, parr2011ll}. 
By defining the grammar rules in the 
grammar definition file, developers can 
implement a parser with high runtime efficiency and minimal grammar rule ambiguities.
A more detailed example of grammar definition 
will be elaborated in~\autoref{ss:design-generation}.
The primary purpose of the parser generator is to verify whether 
the input stream matches the defined syntax rules.
If matched, the parser enables the developer to 
transform the raw input into application's internal structure.
%

Most of the DBMSs known today use parser generators to define their SQL grammar 
rules~\cite{mysql-parser, mariadb-parser, cockroachdb-parser, tidb-parser, 
sqlite-parser, postgres-parser}. 
For example, \mysql, \mariadb and \postgres use \cc{bison} as their parser
generator~\cite{mysql-parser, postgres-parser, mariadb-parser}. 
\cockroachdb and \tidb use a special \cc{GoLang} implemented counterpart of the \cc{yacc}
parser generator~\cite{cockroachdb-parser, tidb-parser, goyacc}.
The \cc{goyacc} used by \cockroachdb and \tidb shares a grammar define notation 
similar to the \cc{bison} one, with both closely aligned with the standard \cc{BNF}.
\sqlite uses a custom parser generator called \cc{Lemon}, which was invented
by the same author who originally developed \sqlite~\cite{sqlite-parser}. 
It adds in a few improvements based on the grammar notations of \cc{bison} or \cc{yacc}, 
but overall maintains a similar grammar definition form.
%
%
In summary, 
the vast usage of parser generators in DBMSs allows us to observe the ground truth grammar 
rules dedicated to different DBMS softwares, and offers us a uniform interface to easily 
leverage these pre-defined grammars for the DBMSs' fuzzing purpose.

\section{Design of \sys}
\label{s:design}

\begin{figure}[t]
  \centering
  \includegraphics[width=.45\textwidth]{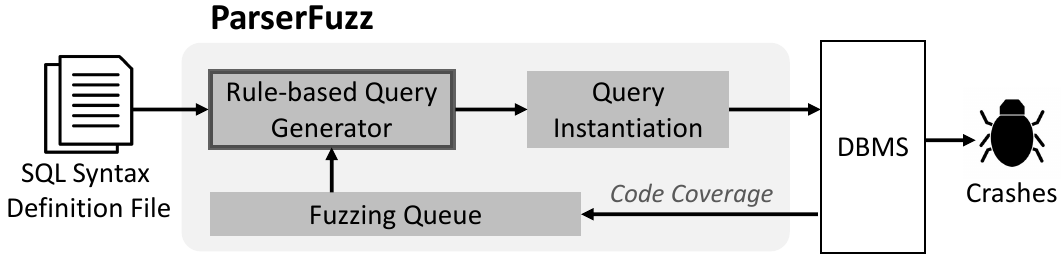}
  \caption{\textbf{Overview of \sys.} 
  It automatically extracts SQL features
  from DBMS-specific syntax definition files, 
  and utilizes code coverage to select promising rules
  for query generation.}
  \label{f:parserfuzz-overview}
\end{figure}

To generate diverse sets of queries that explore all possible 
grammar rules, we introduce parser grammar rule-based query generation.
The query generation doesn't rely on any prior query seeds, 
but instead automatically constructs query statements directly
based on the grammar rules defined for the parser generator.
Additionally, by traversing all the parser rules defined in the
parse generator, we can explore all the possible syntax features tailored 
for the DBMS. 
To steer the fuzzer towards combining different interesting syntax features,
we incorporate code coverage feedback to guide the query mutation.
%
The combination of parser rule-based query generation and code coverage
feedback enables \sys to generate diverse and complex queries that can
effectively find DBMS memory corruption bugs.

\PP{System overview.}
\autoref{f:parserfuzz-overview} shows an overview of our tool \sys.
\sys is the first tool that directly generates SQL query statements based on 
DBMS SQL grammar definition file, and it guarantees exploring all the 
grammar rules defined in the DBMSs.
Unlike previous fuzzers, \sys doesn't require any queries as fuzzing seeds. 
Instead, all queries are produced directly from the parser rule-based 
query generator~\autoref{ss:design-generation}. 
These generated query are then populated with SQL arguments using the validity-oriented 
instantiation algorithm~\autoref{s:impl}. 
All the produced queries are executed in the DBMS afterwards. 
In the current design, \sys produces SQL query sequences according to the following principle:
it firstly generates 3 \cc{CREATE TABLE} statements, 
followed by 3 \cc{INSERT} statements, 2 \cc{CREATE INDEX} statements, 10 randomly generated statements 
regardless of the statement types, and concludes with
10 \cc{SELECT} statements.
\sys will clean up the database contents after one query sequence finishes its 
execution.
If one query statement activates a new code branch from the DBMS,
\sys will save the statement's syntax tree to the fuzzing queue. 
In the next round of generating new statement, \sys offers 50\% chances to 
mutate the queries from the fuzzing queue~\autoref{s:feedback}. 
If any segmentation fault or assertion failure are detected 
during execution, we will save the complete bug-triggering query sequence for 
further analysis.

\begin{figure}[t]
\begin{lstlisting}[style=onlycommenthl, label={l:examplebnf},caption={\textbf{Example grammar definition rules for \cc{Bison}}}]
simple_select:
  SELECT_SYM SCONST 
          {/*extra developer defined logic...*/}
| SELECT_SYM distinct_clause target_list from_clause 
          {/*extra developer defined logic...*/}
| SELECT_SYM distinct_clause target_list from_clause 
        where_clause 
          {/*extra developer defined logic...*/}
;

distinct_clause:
  DISTINCT_SYM {/*extra developer defined logic...*/}
;

target_list:
  target_elem {/*extra developer defined logic...*/}
| target_list ',' target_elem 
        {/*extra developer defined logic...*/}
;

/* other grammar rules ... */
\end{lstlisting}
\end{figure}

\begin{figure}[t]
\begin{lstlisting}[language=c++,label={l:examplelexer},caption={\textbf{Example keyword mappings from the lexer}}]
static const SYMBOL symbols[] = {
  /* SQL keywords (by alphabetical order) */
  {SYM("\&\&", AND_AND_SYM)},
  {SYM("<", LT)},
  {SYM("<=", LE)},
  //...
  {SYM("SELECT", SELECT_SYM)},
  {SYM("DISTINCT", DISTINCT_SYM)},
  //...
};
\end{lstlisting}
\end{figure}

\begin{figure}[t]
\begin{lstlisting}[language=SQL,label={l:exampleparsedquery},caption={\textbf{Querie examples} matching the syntax rules
  in \autoref{l:examplebnf}}]
  -- match the first rule from simple_select
  SELECT 'abc'; 
  -- match the second rule from simple_select
  SELECT DISTINCT c1 FROM t0;   
  -- match the third rule from simple_select
  SELECT DISTINCT c1 FROM t0 WHERE TRUE;
\end{lstlisting}
\end{figure}

\subsection{Parser Rule-based Query Generation}
\label{ss:design-generation}

Exploring the vast and rarely tested SQL features from the DBMS is key
to find more interesting bugs during DBMS fuzzing.
The DBMS SQL rule definition file defines the complete set of the
grammar rules permitted for the SQL dialect. 
%
\autoref{l:examplebnf} shows an example of SQL grammar rule definition.
This grammar notation is adopted by parser generators such as \cc{yacc}, \cc{bison}
and \cc{go-yacc}, and is further used by \mysql, \cockroachdb and \tidb. 
By convention, all lowercase symbols in~\autoref{l:examplebnf} represent non-terminal keywords. 
A non-terminal keyword means the current keyword can be further extended by sub-rules. 
%
For instance, the first line in~\autoref{l:examplebnf} defines the top non-terminal keyword \cc{simple_select}.
Line 2-9 define the sub-rules that can be expanded by the keyword \cc{simple_select}.
Additionally, all uppercase symbols in~\autoref{l:examplebnf}
represent terminal keywords. 
A terminal keyword is one without any attached sub-rule.
They can directly map to a token in the input stream.
\autoref{l:examplelexer} shows an example token mapping.
Alternatively, a terminal keyword can represent arguments 
such as table and column names or as constant values
such as string literals.
The \cc{SCONST} keyword used in~\autoref{l:examplelexer} is an example
for representing any single quoted string constant in the SQL language.
The parser grammar is free to define its grammar entry from 
any non-terminal keyword. 
The parser will recursively match the non-terminal keywords' grammar rules to verify
the grammar correctness.
A successfully parsed query traverses the defined grammar rules, 
constructing a tree-like structure where every tree node represents
a token from the query stream.
%
For instance, if we assume the top keyword \cc{simple_select} 
is the entry point from the grammar syntax definition,
\autoref{l:exampleparsedquery} shows some example queries that can be successfully parsed
by these grammar rules.

Although the main goal of the SQL parser is to
verify SQL inputs, the grammar rules can be easily 
tuned to serve other purposes.
Most interestingly, instead of matching grammar rules from the inputs,
it is feasible to generate query statements from the grammar definition 
by randomly choosing which grammar rule to apply.
Specifically, starting from the grammar entry point, we can randomly select
a grammar rule to commit every time we need to resolve a non-terminal 
keyword. 
By choosing a different set of grammar rules to apply when traversing the syntax,
we can generate various forms of queries for testing.
Moreover, we can easily ensure the
syntax correctness of the generated query because it still adheres to the defined syntax
rules.
Furthermore, to provide the generated queries with semantic meanings, we label all the
SQL arguments in the grammar rules, such as table 
names and column names, addition with constant values 
such as integers and string literals.
These SQL arguments and constant values will later be passed to the validity-oriented 
instantiation algorithm to fill in the concrete values~\autoref{s:impl}.

Because the grammar logics defined for the parser generator serves as 
the front-end of the DBMS processing pipeline, the logic represents the ground 
truth of the grammar rules a DBMS can support.
More importantly, the grammar definition file includes all the grammar features that
the DBMS currently supports.
By thoroughly scanning all the grammar rules defined in the grammar file, 
we can generate queries that saturate the interesting syntax features dedicated 
to the DBMS softwares.
Additionally, the grammar rule exploration does not rely on any prior input corpus,
reducing the burden for DBMS testers to gather an input corpus that covers all of
interesting features they care about.

\begin{figure}[t]
\begin{lstlisting}[style=onlycommenthl, label={l:recursiverule},caption={\textbf{A keyword that recursively references itself}}]
a_expr:
  a_expr '+' a_expr
| a_expr AT_SYM TIME_SYM ZONE_SYM a_expr
| a_expr COLLATE_SYM any_name
| /* other grammar rules ... */
;
\end{lstlisting}
\end{figure}

\begin{figure}[t]
  \begin{lstlisting}[style=onlycommenthl, label={l:simplerule},caption={\textbf{Example simple rule}}]
natural_join_type:
  NATURAL_SYM opt_inner JOIN_SYM /* simple rule */
;

opt_inner:
  %empty    /* syntax termination */
| INNER_SYM /* syntax termination */
;
\end{lstlisting}
\end{figure}

\begin{figure}[t]
\begin{lstlisting}[style=onlycommenthl, label={l:complexrule},caption={\textbf{Example complex rule}}]
subquery: '(' simple_select ')'  ; /* complex rule */
\end{lstlisting}
\end{figure}

\begin{figure}[t]
\begin{lstlisting}[style=onlycommenthl, label={l:ruleprioritization},caption={\textbf{Rule prioritization to resolve path explostion}}]
table_reference:
  table_factor   /* normal rule, preferred */
| joined_table   /* complex rule */
;

table_factor:
  table_name     /* simple rule, preferred */
| table_function /* complex rule */
;

joined_table:
  /* complex rule */
  table_reference inner_join_type table_reference 
  /* complex rule */
| table_reference outer_join_type table_reference
;
\end{lstlisting}
\end{figure}

\subsection{Path Explosion due to Recursive Keyword}

Although the BNF notation isn't as complex as programming languages such as \cc{C/C++}, 
we also face similar issues with program static analysis 
when generating new queries from the grammar rules.
The most notable issue is the grammar path explosion problem.
\autoref{l:recursiverule} demonstrates an example that could lead
to grammar path explosion.
Most of the grammar rules defined under the \cc{a_expr} keyword reference the top keyword
\cc{a_expr}, which creates a loop in the grammar rule that continuously expands the grammar
tree. 
Given most of the grammar rules defined in \autoref{l:recursiverule} have recursive
references, and some rules such as \cc{a_expr `+' a_expr} even reference the top keyword twice,
the query generation algorithm might get trapped in this recursive keyword and never finds
an exit. 
Because \cc{a_expr} is an important grammar feature
in the DBMS, addressing this grammar path explosion problem becomes one important task
to tackle.

To address this grammar path explosion problem, \sys scans through the
query grammar files and classifies all the grammar rules into three categories: \cc{simple rule},
\cc{normal rule} and \cc{complex rule}.
Simple rules are any grammar rules that guaranteed to terminate in 2 iterations.
\autoref{l:simplerule} shows one example. 
The rule defined under the \cc{natural_join_type} keyword contains two terminal keywords: 
\cc{NATURAL_SYM}, \cc{JOIN_SYM} and one non-terminal keyword \cc{opt_inner}.
However, all grammar rules defined for \cc{opt_inner} definitively lead to rule
termination, \ie, they could be either \cc{empty} or \cc{INNER_SYM}. 
In this case, by choosing the grammar rule \cc{NATURAL_SYM opt_inner JOIN_SYM}, \sys ensures
the grammar rule will cease expansion in 2 iterations. 
\sys therefore labels this grammar rule as simple rule.
Complex rules are syntax definitions that lead to recursive keywords or more intricate grammar expressions.
Apart from the aforementioned case in \autoref{l:recursiverule}, \autoref{l:complexrule} illustrates another complex rule 
example, which expands into an entire sub-select statement within the query expression, significantly complicating the syntax 
tree.
\sys automatically labels all grammar rules that recursively reference keywords as complex rules.
User can also manually label certain complex grammar rules, like the one shown in \autoref{l:complexrule}. 
Normal rules are all the other grammar rules excluding simple rules and complex rules.
They represent an uncertain status of the current syntax rule, where the defined syntax will neither lead
to immediate termination, nor lead to path explosion.
User can manually define certain rules as either simple rules or complex rules to guide the query generation, 
and \sys handles the normal rule sampling automatically.

After categorizing the grammar rules, \sys uses the categorization information to guide the query generation.
In the initial stages of the query generation, \ie, within a shallow syntax rule iteration depth, \sys 
freely selects grammar rules to commit based on a Multi-Armed Bandit (MAB) solver 
described in \autoref{s:feedback}.
Once the query generation reaches a specific depth, \sys prioritizes generating 
$simple\_rule > normal\_rule > complex\_rule$.
\autoref{l:ruleprioritization} illustrates one example that uses rule prioritization 
to early-exit the query generation.
The \cc{table_reference} keyword in \autoref{l:ruleprioritization} contains two grammar rules.
The first rule is a \cc{normal rule}, because the grammar defined in \cc{table_factor} could lead to 
both simple and complex rules.
Conversely, the second rule from \cc{table_reference} is a complex rule because all the grammar features
specified in \cc{join_table} recursively reference the top keyword \cc{table_reference}. 
Therefore, once reaching specific depth, \sys prioritizes 
normal rule, \cc{table_factor}, when resolving \cc{table_reference}. 
It then commits to the rule \cc{table_name}, which is a simple rule defined after \cc{table_factor}, 
and it concludes the current syntax tree path.
This categorization-based parser rule prioritization significantly enhances the success rate of generating
valid SQL queries,
because it effectively avoids the path explosion problem during query generation.

\begin{figure}[t]
\begin{lstlisting}[language=SQL,label={l:examplemutation},caption={\textbf{Example mutation on saved query}}]
/* interesting query saved in the fuzzing queue */
SELECT * FROM t0 INNER JOIN t1 ON t0.c1 = t1.c1;
/* the query are mutated to the following */
SELECT * FROM t0 OUTER JOIN t1 ON t0.c1 = t1.c1;
\end{lstlisting}
\end{figure}

\begin{figure}[t]
\begin{lstlisting}[label={l:examplemutationrules},caption={\textbf{Grammar rules for the example in \autoref{l:examplemutation}}}]
join_table:
  table_reference INNER_SYM JOIN_SYM table_reference
    {/*extra user defined logic...*/}
| table_reference OUTER_SYM JOIN_SYM table_reference
| /* other grammar rules ... */
;

\end{lstlisting}
\end{figure}

\begin{figure}[t]
\begin{lstlisting}[label={l:examplecodecoveragemab},caption={\textbf{Rule prioritization based on code coverage}}]
a_expr:
  /* other grammar rules ... */
| STRING_LITERAL /* new coverage 1 time */
| a_expr COLLATE_SYM any_name /* new coverage 3 times */
| /* other grammar rules ... */ 
;
\end{lstlisting}
\end{figure}

\subsection{Query Mutation with Coverage Feedback}
\label{s:feedback}

Lessons learned from previous grey-box fuzzing tools suggest that code coverage feedback
is effective in leading fuzzers into finding more bugs.
%
After executing a query, \sys gathers the code branches
triggered in the DBMS.
If a query triggers a new code branch, \sys collects the syntax tree that constructs the current executed 
query into its fuzzing queue. 
Because each query acts as an independent statement in the \sys generation process, 
\sys does not save concrete values such as arguments or constant values into the
syntax tree.
On the next query generation, \sys offers 50\% chances to mutate 
the syntax trees from the fuzzing queue instead of generating a new one.
Furthermore, \sys has the autonomy to choose any tree nodes from the syntax tree to 
start its query mutation. 
To mutate, \sys refers to the SQL syntax definition rules, starts 
generating new query segments from the 
randomly chosen query mutation point, and ultimately 
replaces the original query segments with the newly produced one.
\autoref{l:examplemutation} shows one such mutation example. 
Assuming the original query from~\autoref{l:examplemutation} triggers a new code coverage 
from the DBMS, so it is stored in the fuzzing queue. 
During the mutation, \sys selects the \cc{join_table} node to mutate.
The corresponding grammar rule definition is shown in~\autoref{l:examplemutationrules},
and it covers the query segment \cc{t0 INNER JOIN t1}. 
\sys then randomly selects another parser rule that defined under \cc{join_table}, and commits 
to the second rule in~\autoref{l:examplemutationrules}.
The original query from \autoref{l:examplemutation} is ultimately mutated to the form with \cc{OUTER JOIN}.

In addition of using the code coverage feedback to save queries for further mutations,
\sys also uses the code coverage feedback to prioritize more 
feature-rich grammar rules that might
lead to interesting DBMS behaviors.
Certain grammar rules are doomed to be more feature-rich and interesting than others.
For example, in the case of~\autoref{l:examplecodecoveragemab}, 
the first rule on line 3 resolves a constant string, 
making it less interesting than the second rule defined in line 4, 
where the latter rewrites the default collation 
for the expression's return value. 
Because the second rule can trigger more unique code coverage compared to the
constant string resolving during the fuzzing samples, 
\sys should favor the \cc{COLLATION} 
rule when formulating a new SQL statement.
In \sys, we model the grammar rule commitment as an Multi-Arm Bandit (MAB) problem,
where \sys plays a game every time it needs to opt for a grammar rule
to commit. 
If the generated query triggers a new code coverage, all the grammar rules used to construct the test query
receive a reward.
The fuzzer's objective is to generate queries that maximize the code coverage(reward).
However, the information of how much code coverage is achievable remains limited or unknown 
at the time of committing to the 
grammar rules.
Thus, \sys needs to conduct an optimized strategy that explores all the different grammar possibilities
while also
favoring selecting the grammars that induce more captivating outcomes.
In \sys, we use the $\epsilon$-Greedy algorithm to address this MAB problem.
The DBMS testers can predetermine an $\epsilon$ value,
which represents the possibility of directly selecting the grammar rule with the highest known reward. 
Conversely, with a $1 - \epsilon$ probability, \sys randomly selects any rules defined 
to maximize the exploration.
By default, the $\epsilon$ value for \sys is set to be 0.5.
But users can adjust the $\epsilon$ value to suit their needs.

\section{Implementation}
\label{s:impl}

We implemented \sys
based on the logic-bug detection tool
\sqlright~\cite{liang2022sqlright}.
The query-instantiation logic
of \sqlright can handle a variety of 
semantic situations.
We removed the oracle interfaces from \sqlright and
expanded the fuzzer to accept any valid SQL statements. 
Moreover,
we updated \sqlright's query mutation logic,
changing the tool from
a mutation-based fuzzer that relies on input corpus
to a generation-based fuzzer
that can generate queries by parsing grammar definition files.
Here, we present more implementation details for the users that are interested in \sys.

\PP{Rule-based query generator.} 
This generator is built
based on a prototype developed by \cc{Cockroach Labs},
the developer of \cockroachdb.
The prototype is named \cc{RSG},
standing for Random Statement Generator~\cite{rsgcockroachdbsrc}.
However,
this prototype supports only
a limited number of features
from the BNF parser notation. 
For example,
the prototype does not recognize notation \cc{\%prec},
which is reserved for the parser generator usage.
Moreover, the prototype struggles with complex syntax rules, particularly with recursive keywords.
As a result, this prototype was primarily designed to 
generate simple and short DEMO query statements.
The grammar definition files provided to the prototype rarely exceed 15 lines,
making the prototype unsuitable for parsing the comprehensively grammar rules 
designed for \cockroachdb.
We enhanced the prototype's capabilities to accommodate more complex grammar definition rules, which includes 
adding full support for \cc{go-yacc} grammar files used in \cockroachdb and \tidb, \cc{bison} grammar files 
used in \mysql and \mariadb, and even \cc{Lemon} grammar file specifically designed for \sqlite.
The rule-based query generator in \sys can be further extended to support other grammar definition formats in the
future, including the support for modern parser generation tool \cc{ANTLR}.

\definecolor{light-gray}{gray}{0.88}

\begin{table*}[!ht]
\small
\footnotesize
\centering
\setlength{\tabcolsep}{4pt}

\begin{tabular}{@{}ccllccccc@{}}
  \toprule

  \textbf{DBMS}  &  \textbf{ID} &  \textbf{Description} & \textbf{Status} & \textbf{\squirrel} & \textbf{\sqlsmith} & \textbf{\cockroachdbsqlsmith} & \textbf{\gosqlsmith} & \textbf{\sqlancerqpg} \\ 
  \midrule


  \multirow{2}{*}{\STAB{\rotatebox[origin=c]{90}{\sqlite}}}
  & 1  & \cc{RETURNNING} from ill-formed \cc{VIEW}                     & fixed (84417bbd)        & \no   & \yes  & -     & -     & \no   \\
  & 2  & Unexpected exposed debug function                             & fixed (62114711)        & \yes  & \yes  & -     & -     & \no   \\
  & 3  & Incorrect byte code conversion                                & fixed (8f637aae)        & \no   & \yes  & -     & -     & \no   \\

  \hline

  \multirow{8}{*}{\STAB{\rotatebox[origin=c]{90}{\mysql}}}
  & 4  & Incorrect sorting optimization                                & fixed (version 8.0.34)  & \no   & -     & -     & -     & -     \\
  & 5  & Incorrect partition condition handling                        & confirmed               & \no   & -     & -     & -     & -     \\
  & 6  & Incorrect partition condition handling                        & confirmed               & \no   & -     & -     & -     & -     \\
  & 7  & Incorrect \cc{REGEXP} expression handling                     & confirmed               & \no   & -     & -     & -     & -     \\
  & 8  & \cc{TEMP TABLE} created with ill-formed function index        & confirmed               & \no   & -     & -     & -     & -     \\
  & 9  & Incorrect \cc{CHECK} condition handling in \cc{CREATE TABLE}  & confirmed               & \no   & -     & -     & -     & -     \\
  & 10 & Incorrect charset conversion                                  & confirmed               & \no   & -     & -     & -     & -     \\
  & 11 & Incorrect subquery expression handling                        & confirmed               & \no   & -     & -     & -     & -     \\

  \hline

  \multirow{7}{*}{\STAB{\rotatebox[origin=c]{90}{\cockroachdb}}}
  & 12 & Incorrect temp disk storage internal value                    & fixed (1ee803ee)        & -     & -     & \yes  & -     & \no   \\
  & 13 & Duplicate \cc{PRIMARY KEY}                                    & fixed (3a3123d9)        & -     & -     & \no   & -     & \no   \\
  & 14 & Missing name resolver to constraint validator                 & fixed (0c58a08d)        & -     & -     & \yes  & -     & \no   \\
  & 15 & Incorrect data type conversion                                & fixed (5cb5d1da)        & -     & -     & \yes  & -     & \yes  \\
  & 16 & Incorrect default expression typing and backfill              & fixed (51005e41)        & -     & -     & \yes  & -     & \no   \\
  & 17 & Out-of-bounds from tuple handling                             & fixed (58ec9687)        & -     & -     & \yes  & -     & \yes  \\
  & 18 & Large number as hidden constants                              & fixed (318e352e)        & -     & -     & \yes  & -     & \yes  \\

  \hline

  \multirow{7}{*}{\STAB{\rotatebox[origin=c]{90}{\tidb}}}
  & 19 & Incorrect query parsing logic                                 & fixed (8f308ecb)        & -     & -     & -     & \no   & \no   \\
  & 20 & Compare subquery in \cc{SHOW}                                 & confirmed               & -     & -     & -     & \no   & \no   \\
  & 21 & Index out of bound access in \cc{EXPLAIN}                     & fixed (762432b6)        & -     & -     & -     & \no   & \no   \\
  & 22 & DML panic when CTE exists                                     & fixed (25764bc8)        & -     & -     & -     & \no   & \no   \\
  & 23 & Incorrect expression rewriter optimization                    & confirmed               & -     & -     & -     & \no   & \yes  \\
  & 24 & Recovery non-existing jobs                                    & confirmed               & -     & -     & -     & \no   & \no   \\
  & 25 & Incorrect handling for partial aggregation                    & confirmed               & -     & -     & -     & \no   & \no   \\

  \hline

  \multirow{4}{*}{\STAB{\rotatebox[origin=c]{90}{\mariadb}}}
  & 26 & Incorrect partition condition handling                        & confirmed               & \no   & -     & -     & -     & -     \\
  & 27 & Incorrect check condition handling in \cc{CREATE TABLE}       & fixed (8adb6107)        & \yes  & -     & -     & -     & -     \\
  & 28 & Incorrect remove record without match in \cc{DELETE}          & confirmed               & \no   & -     & -     & -     & -     \\
  & 29 & Incorrect sub-select optimization                             & confirmed               & \yes  & -     & -     & -     & -     \\

\bottomrule
    
\end{tabular}
  \caption{\textbf{New Segmentation Faults detected by \sys}. \label{t:all-segfaults}
  \sys detects \totalbugs bugs in total, 
  including \crashes crashes and \assertbugs assertion failures. 
  The \squirrel, \sqlsmith, \cockroachdbsqlsmith, \gosqlsmith and \sqlancerqpg columns 
  represent whether the referenced tools can theoretically detect 
  the mentioned bug, `\yes' means `Yes', `\no' states `No' and `-' means the tool is not
  applicable to the target DBMS.
  }
\end{table*}

\begin{table*}[!ht]
\small
\footnotesize
\centering
\setlength{\tabcolsep}{2.2pt}
  
\begin{tabular}{@{}ccllccccc@{}}
  \toprule

  \textbf{DBMS}  &  \textbf{ID} &  \textbf{Description} & \textbf{Status} & \textbf{\squirrel} & \textbf{\sqlsmith} & \textbf{\cockroachdbsqlsmith} & \textbf{\gosqlsmith} & \textbf{\sqlancerqpg} \\ 

  \midrule


  \multirow{2}{*}{\STAB{\rotatebox[origin=c]{90}{\sqlite}}}
  & 1  & `pExpr->affExpr==OE_Rollback ...'                                 & fixed (e9543911)       & \no   & \no   & -     & -     & \no    \\
  & 2  & sqlite3_result_blob, `n>=0'                                       & fixed (ab3331f4)       & \no   & \yes  & -     & -     & \no    \\
  & 3  & `sqlite3VdbeMemValidStrRep(pVal)'                                 & fixed (3e2da8a7)       & \no   & \no   & -     & -     & \no    \\

  \hline

  \multirow{19}{*}{\STAB{\rotatebox[origin=c]{90}{\mysql}}}
  & 4  & `escape_arg != nullptr'                                           & fixed (version 8.2.0)  & \no   & -     & -     & -     & -      \\
  & 5  & `m_alter_info->requested_lock'                                    & confirmed              & \no   & -     & -     & -     & -      \\
  & 6  & `has_error == thd->get_stmt_da()->is_error()'                     & confirmed              & \no   & -     & -     & -     & -      \\
  & 7  & check_set_user_id_priv, `0'                                       & fixed (version 8.0.35) & \yes  & -     & -     & -     & -      \\
  & 8  & `is_prepared() \&\& !is_optimized()'                              & confirmed              & \no   & -     & -     & -     & -      \\
  & 9  & MoveCompositeIteratorsFromTablePath, `false'                      & confirmed              & \no   & -     & -     & -     & -      \\
  & 10 & `!thd->lex->is_exec_started()'                                    & confirmed              & \no   & -     & -     & -     & -      \\
  & 11 & `!sl->order_list.first'                                           & confirmed              & \yes  & -     & -     & -     & -      \\
  & 12 & `m_return_field_def.auto_flags == Field::NONE'                    & confirmed              & \no   & -     & -     & -     & -      \\
  & 13 & `m_relaylog_file_reader.position() == m_rli->...'                 & confirmed              & \no   & -     & -     & -     & -      \\
  & 14 & `ha_alter_info->handler_flags ...'                                & confirmed              & \yes  & -     & -     & -     & -      \\
  & 15 & `!thd->lex->is_exec_started() || thd->lex ...'                    & confirmed              & \no   & -     & -     & -     & -      \\
  & 16 & `!thd->in_sub_stmt'                                               & confirmed              & \no   & -     & -     & -     & -      \\
  & 17 & `is_prepared()'                                                   & confirmed              & \no   & -     & -     & -     & -      \\
  & 18 & `thd->is_error()'                                                 & fixed (79eae6a2)       & \no   & -     & -     & -     & -      \\
  & 19 & `inited == NONE || table->open_by_handler'                        & confirmed              & \yes  & -     & -     & -     & -      \\
  & 20 & `is_nullable()'                                                   & confirmed              & \no   & -     & -     & -     & -      \\
  & 21 & `!is_set()'                                                       & confirmed              & \no   & -     & -     & -     & -      \\
  & 22 & `m_deque == other.m_deque'                                        & confirmed              & \yes  & -     & -     & -     & -      \\

  \hline

  \multirow{30}{*}{\STAB{\rotatebox[origin=c]{90}{\cockroachdb}}}
  & 23 & unsupported comparison operator                                   & fixed (7b473a8f)       & -     & -     & \yes  & -     & \yes   \\
  & 24 & input to ArrayFlatten should be uncorrelated                      & confirmed              & -     & -     & \yes  & -     & \yes   \\
  & 25 & an empty end boundary must be inclusive                           & confirmed              & -     & -     & \no   & -     & \yes   \\
  & 26 & runtime error: index out of range                                 & confirmed              & -     & -     & \no   & -     & \yes   \\
  & 27 & unexpected error from the vectorized engine                       & fixed (8d1865fd)       & -     & -     & \no   & -     & \yes   \\
  & 28 & tuple length mismatch                                             & confirmed              & -     & -     & \no   & -     & \no    \\
  & 29 & use of crdb_internal_vtable_pk column not allowed                 & fixed (5cc456bb)       & -     & -     & \no   & -     & \no    \\
  & 30 & top-level relational expression cannot have outer columns         & confirmed              & -     & -     & \no   & -     & \no    \\
  & 31 & cannot map variable 7 to an indexed var                           & confirmed              & -     & -     & \no   & -     & \no    \\
  & 32 & expected *DString, found tree.dNull                               & fixed (6eabc2f3)       & -     & -     & \no   & -     & \no    \\
  & 33 & invalid memory address or nil pointer dereference                 & fixed (b4d5b0b8)       & -     & -     & \yes  & -     & \no    \\
  & 34 & aggregate function is not allowed in this context                 & fixed (1c8dd156)       & -     & -     & \no   & -     & \no    \\
  & 35 & invalid memory address or nil pointer dereference                 & fixed (de8a3c77)       & -     & -     & \yes  & -     & \no    \\
  & 36 & expected subquery to be lazily planned as routines                & fixed (9f319ddb)       & -     & -     & \yes  & -     & \no    \\
  & 37 & tuple contents and labels must be of same length: [], [alias_0]   & confirmed              & -     & -     & \no   & -     & \no    \\
  & 38 & unhandled type *tree.RangeCond                                    & fixed (0d647800)       & -     & -     & \no   & -     & \no    \\
  & 39 & referenced descriptor ID 1: descriptor not found                  & confirmed              & -     & -     & \no   & -     & \no    \\
  & 40 & invalid datum type given: inet, expected int                      & fixed (ff87db04)       & -     & -     & \no   & -     & \no    \\
  & 41 & unexpected statement: *tree.SetTracing                            & confirmed              & -     & -     & \no   & -     & \no    \\
  & 42 & cannot overwrite distribution ...                                 & confirmed              & -     & -     & \no   & -     & \no    \\
  & 43 & no output column equivalent to 6                                  & fixed (b9b8da67)       & -     & -     & \yes  & -     & \yes   \\
  & 44 & index out of range [0] with length 0 (in function handling)       & confirmed              & -     & -     & \no   & -     & \no    \\
  & 45 & unrecognized relational expression type: alter-table-unsplit-all  & confirmed              & -     & -     & \no   & -     & \no    \\
  & 46 & schema change PostCommitPhase, index out of range [1]             & fixed (f0dede19)       & -     & -     & \no   & -     & \no    \\
  & 47 & generator functions cannot be evaluated as scalars                & confirmed              & -     & -     & \no   & -     & \no    \\
  & 48 & could not parse "1 sec" as type bool: invalid bool value          & confirmed              & -     & -     & \no   & -     & \yes   \\
  & 49 & SetAnnotation(), index out of range [4] with length 1             & fixed (c6cf5189)       & -     & -     & \no   & -     & \no    \\
  & 50 & locking cannot be used with virtual table                         & confirmed              & -     & -     & \no   & -     & \no    \\
  & 51 & no known encoding type for array                                  & confirmed              & -     & -     & \yes  & -     & \no    \\
  & 52 & zero transaction timestamp in EvalContext                         & confirmed              & -     & -     & \no   & -     & \no    \\

\bottomrule
    
\end{tabular}
  \caption{\textbf{New Assertion Failures detected by \sys}. \label{t:all-asserts}
  \sys detects \totalbugs bugs in total, 
  including \crashes crashes and \assertbugs assertion failures. 
  The \squirrel, \sqlsmith, \cockroachdbsqlsmith, \gosqlsmith and \sqlancerqpg columns 
  represent whether the referenced tools can theoretically detect 
  the mentioned bug, `\yes' means `Yes', `\no' states `No' and `-' means the tool is not
  applicable to the target DBMS.
  \cc{Assertion failure} for \cockroachdb represents the bug that 
  \cockroachdb returns unexpected error. But the bug would 
  not crash the whole \cockroachdb process.
  }

\end{table*}

\PP{DBMS coverage instrumentation} 
We used AFL LLVM mode to instrument the DBMSs that are written in \cc{C} or \cc{C++} languages~\cite{afl}, including
\sqlite, \mysql and \mariadb.
In addition, we enlarged the code coverage map size from \cc{64K} to \cc{256K}. 
However, we couldn't find any existing method to apply branch coverage instrumentation 
for DBMSs implemented in \cc{GoLang}.
Therefore, we modified the line coverage instrumentation from \cc{GoLang} built-in library~\cite{gofuzzdocumentation}, 
and enhanced its capability
to support branch coverage logging.
We then integrated our custom branch coverage feedback logging in \sys when testing \cc{GoLang} implemented DBMSs 
such as \cockroachdb and \tidb.

\section{Evaluation}
\label{s:eval}

We evaluate \sys on five popular open-source DBMSs,
including \sqlite, \mysql, \cockroachdb, \tidb and \mariadb.
The evaluation aims to answer the following questions.

\begin{enumerate}[nolistsep,noitemsep,label=\textbf{Q{\arabic*}.},ref=\textbf{Q\arabic*}]

\item \label{q:bugs} Can \sys detect real-world DBMS bugs?
  
\item \label{q:comparison} Can \sys find more bugs than existing tools?

\item \label{q:feedback} How does code coverage guide the fuzzing process?

\item \label{q:casestudy} How do extra syntax rules contribute to bug finding?

\end{enumerate}

\PP{Experimental setup}
To address \autoref{q:bugs}, we conduct experiments of 
\sys on 5 popular DBMSs, \sqlite, \mysql, \cockroachdb, \tidb and \mariadb,
and gather all the bugs detected in~\autoref{ss:eval-bugs}.
%
%
To answer~\autoref{q:comparison}, we compare \sys to existing state-of-the-arts 
in~\autoref{ss:eval-comparison}.
Due to the diverse SQL dialects in different DBMSs, 
there is no universal DBMS testing tool that covers all the DBMSs we are testing.
Therefore, for each evaluated DBMS,
we select the latest open-source DBMS testing programs 
that are compatible to the DBMS as baselines.
For \sqlite, \mysql and \mariadb, we compare \sys against \squirrel, the most
advanced grey-box mutation-based DBMS fuzzer.
For \cockroachdb and \tidb, we use the official query generation-based testing
tools that are maintained by the DBMS developer groups, \ie, we test the customized SQLsmith 
(\cockroachdbsqlsmith for short)
for \cockroachdb, and test go-sqlsmith (\gosqlsmith for short) for \tidb respectively.
To compare \sys against traditional bit-flips mutation-based fuzzer, we select
\aflpp to test \cc{C/C++} implemented DBMSs and use \libfuzzer to test \cc{GoLang}
implemented ones.
In addition, to understand the memory error detecting capability for DBMS logic bug
detectors, we compare \sys to state-of-the-art logic bug testing tool \sqlancerqpg.
\sqlancerqpg supports testing with \sqlite, \cockroachdb and \tidb, and  
outperforms all other logic bug detectors including \sqlright\cite{ba2023testing,liang2022sqlright}.
We use \norec oracle for \sqlancerqpg when testing with \sqlite and \cockroachdb. 
Because \norec oracle is claimed to be a better performer overall 
compared to \tlp oracle\cite{ba2023testing}.
But we fallback to use \tlp when testing \tidb, because \sqlancerqpg hasn't
supported testing \tidb with \norec oracle yet.
While the most recent DBMS fuzzing tool \dynsql~\cite{jiang2023dynsql} supports testing 6 DBMSs including
\sqlite, \mysql and \mariadb, it is not open-source, so we cannot 
compare our tool to their implementation.
For fuzzing tools that demand input corpus, we use the query libraries 
from the \squirrel repo to serve as the universal input seeds.
To answer~\autoref{q:feedback}, we disable the code coverage feedback from \sys, 
transforming it into a pure random query generation tool, which noted as \sysnocov.
We compare \sysnocov against the full-featured 
\sys in~\autoref{ss:eval-feedback}. 
Finally, we use the bugs detected to demonstrate the contribution
of the diverse syntax elements from \sys~\autoref{ss:eval-syntax}, which answers the question of \autoref{q:casestudy}.

We run all evaluations on an Ubuntu 20.04 system.
The machine comes with two 28-cores Intel(R) Xeon(R) Gold 6348 CPUs and 512 GB memory.
We target the latest release versions of the DBMSs 
at the time when we started the evaluation.
Specifically, we evaluate \sqlite on version \cc{3.41.0}, \mysql on version \cc{8.0.33}, 
\cockroachdb on version \cc{v22.1.10}, \tidb on version \cc{v6.1.7} and \mariadb on version
\cc{11.3}. 

\subsection{DBMS Bugs}
\label{ss:eval-bugs}

Due to the resource limitations, we evaluated different
DBMSs over different testing durations.
We fuzzed \mysql with the longest time frame, which lasted
3 months. 
We further tested \sqlite for 2 months, \cockroachdb for 2 months,
\tidb for 1 month and \mariadb for 3 weeks.
In total, \sys detected \totalbugs bugs from all 5 DBMSs, containing
\crashes segmentation faults and \assertbugs assertion failures.

A bug summary is presented in~\autoref{t:all-segfaults} and~\autoref{t:all-asserts}. 
A segmentation fault indicates a bug that brings down the 
DBMS server process, and forces the DBMS client to exit the ongoing session.
An attacker can effectively exploit a segmentation fault PoC to 
conduct Denial-of-Service 
attack on any online DBMS services.
An Assertion failure from \sqlite and \mysql implies the provided PoC is reproducible
only in debug build of the DBMS.
Although the release build of the DBMS does not trigger the assertion crash, the failed
assertion check implies that the DBMS operates in an ill-formed state. 
An attacker might exploit this ill-formed state to trigger higher impact
exploitation.
An assertion failure from \cockroachdb represents an unexpected runtime error.
The cause can be as fatal as \cc{index out of bounds access}, 
however, \cockroachdb automatically recovers from the error state, and it will
discard the malicious changes and then resume running. 

\subsection {Comparison with Existing Tools}
\label{ss:eval-comparison}

\begin{figure*}[!h]
\begin{minipage}{0.8\textheight}

\centering

\begin{minipage}{\textwidth}
\begin{center}
\begin{subfigure}[c]{\textwidth}
\center{
  \includegraphics[width=0.8\textwidth]{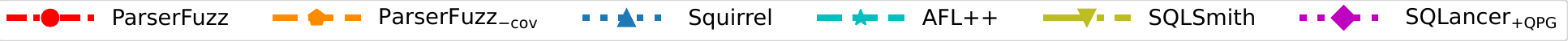}
}
\end{subfigure}
\begin{subfigure}[c]{\textwidth}
\center{
  \includegraphics[width=0.56\textwidth]{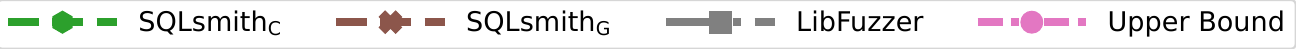}
}
\end{subfigure}
  \end{center}
\end{minipage}

\vspace{2.5mm}

\begin{subfigure}[c]{0.24\textwidth}
  \includegraphics[width=\textwidth]{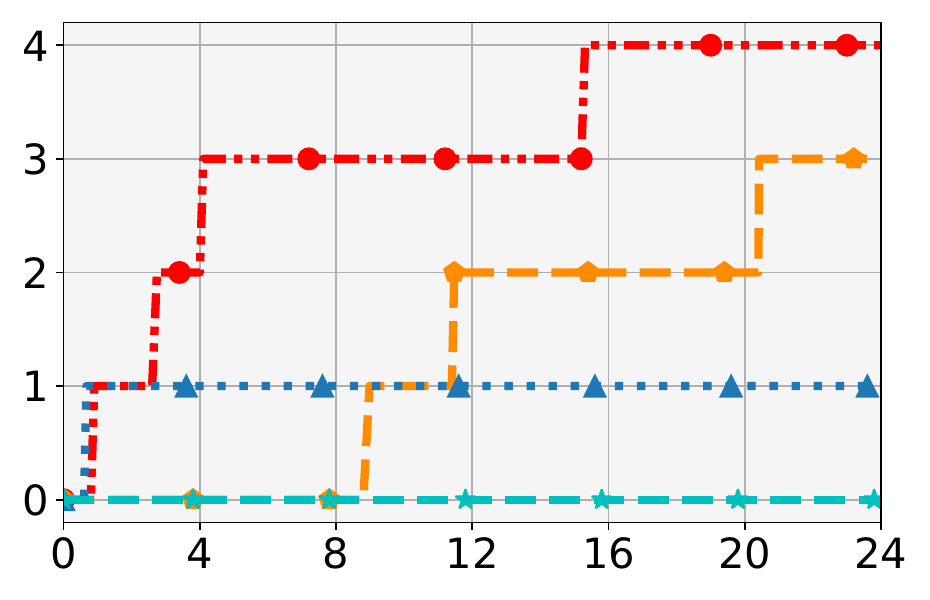}
  \caption{\mysql: detected bugs}
  \label{fig:mysql-bug}
\end{subfigure}
\begin{subfigure}[c]{0.24\textwidth}
  \includegraphics[width=\textwidth]{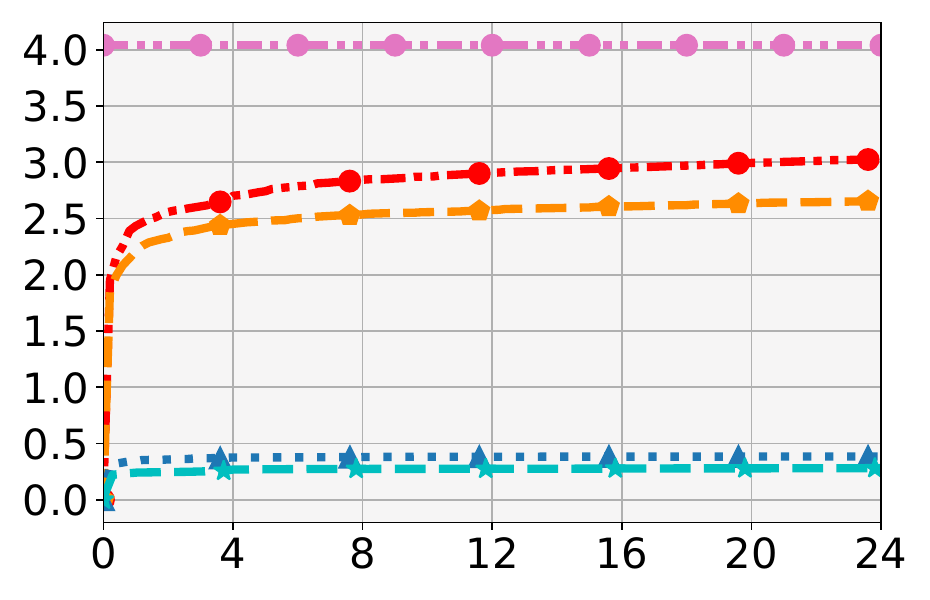}
  \caption{\mysql: grammar edges (k)}
  \label{fig:mysql-gram-size}
\end{subfigure}
\begin{subfigure}[c]{0.24\textwidth}
  \includegraphics[width=\textwidth]{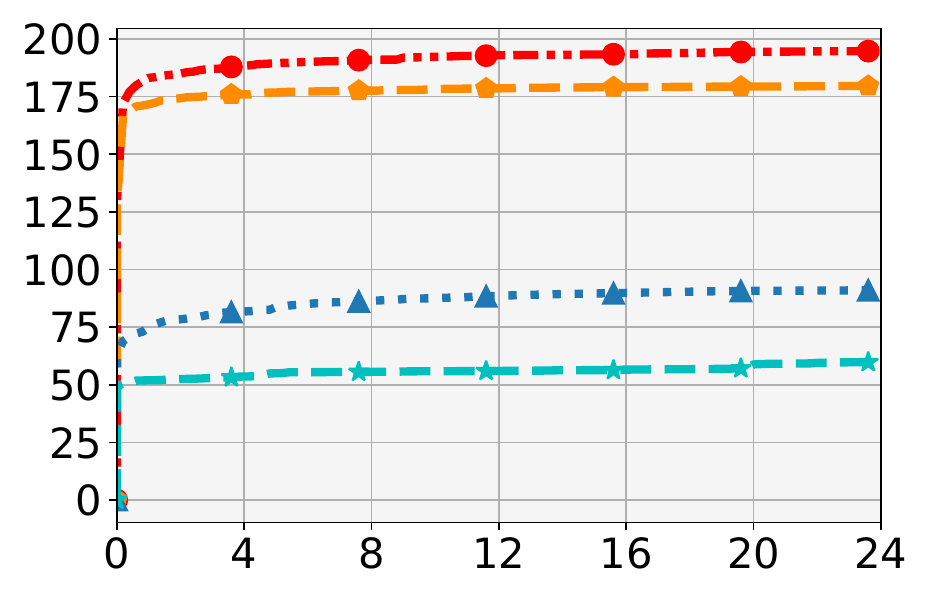}
  \caption{\mysql: code coverage (k)}
  \label{fig:mysql-code-coverage}
\end{subfigure}
\begin{subfigure}[c]{0.24\textwidth}
  \includegraphics[width=\textwidth]{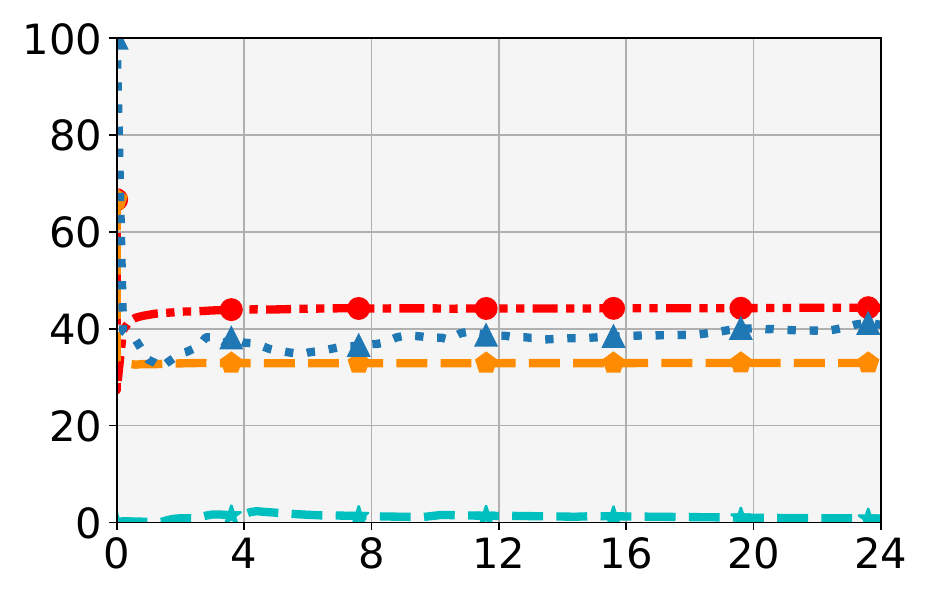}
  \caption{\mysql: query validity (\%)}
  \label{fig:mysql-correct-rate}
\end{subfigure}

\vspace{2.5mm}


\begin{subfigure}[c]{0.24\textwidth}
  \includegraphics[width=\textwidth]{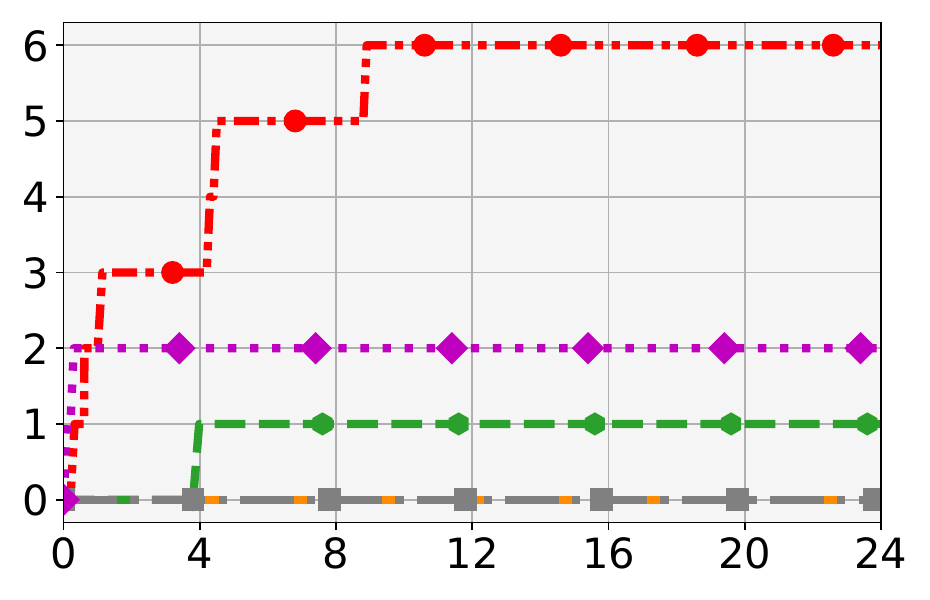}
  \caption{\cockroachdb: detected bugs}
  \label{fig:cockroachdb-bug}
\end{subfigure}
\begin{subfigure}[c]{0.24\textwidth}
  \includegraphics[width=\textwidth]{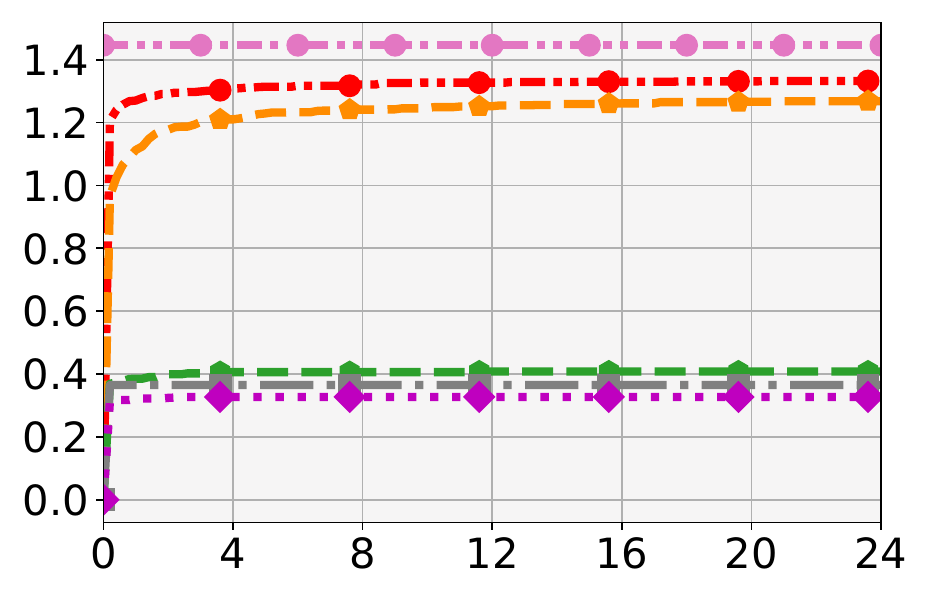}
  \caption{\cockroachdb: grammar edges (k)}
  \label{fig:cockroachdb-gram-size}
\end{subfigure}
\begin{subfigure}[c]{0.24\textwidth}
  \includegraphics[width=\textwidth]{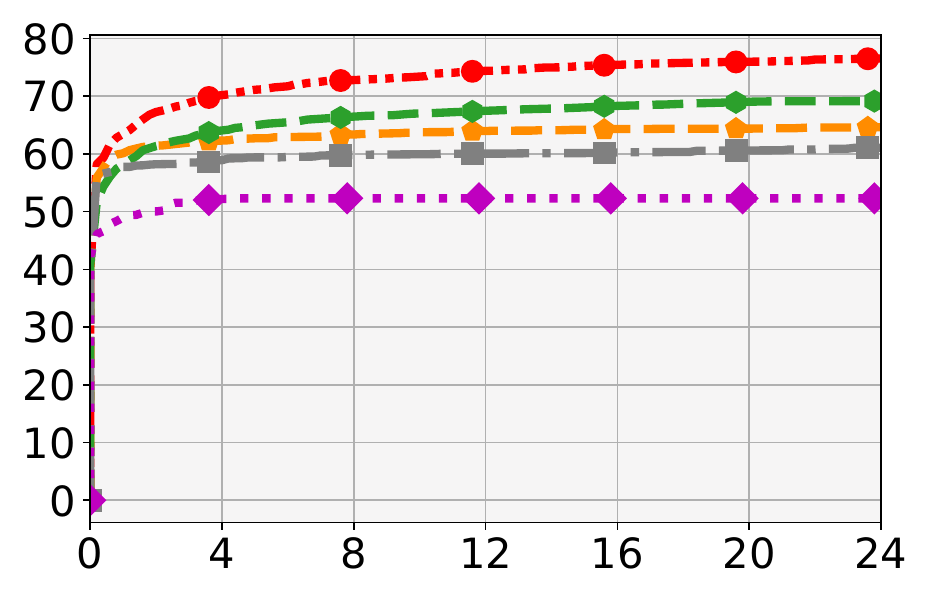}
  \caption{\cockroachdb: code coverage (k)}
  \label{fig:cockroachdb-code-coverage}
\end{subfigure}
\begin{subfigure}[c]{0.24\textwidth}
  \includegraphics[width=\textwidth]{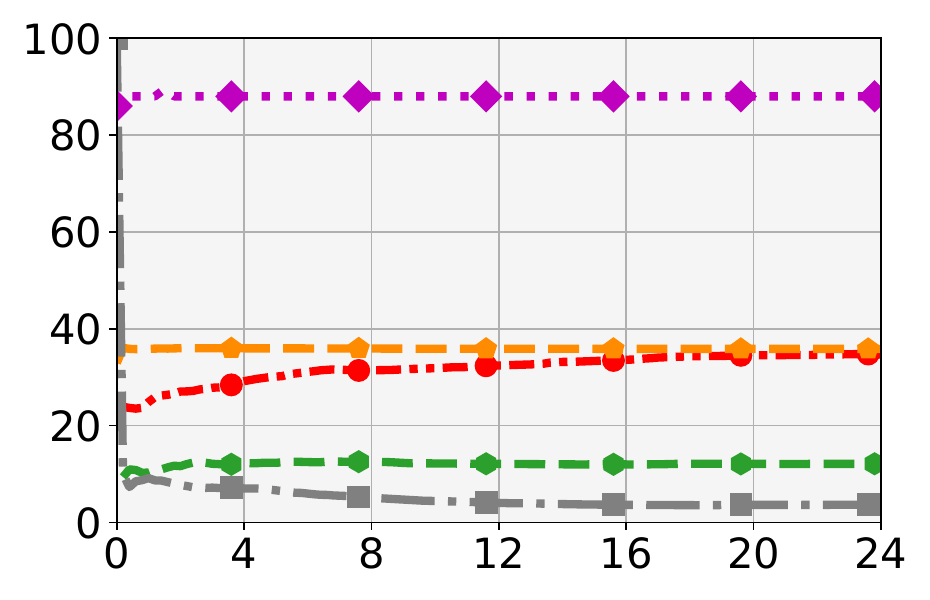}
  \caption{\cockroachdb: query validity (\%)}
  \label{fig:cockroachdb-correct-rate}
\end{subfigure}

\vspace{2.5mm}


\begin{subfigure}[c]{0.24\textwidth}
  \includegraphics[width=\textwidth]{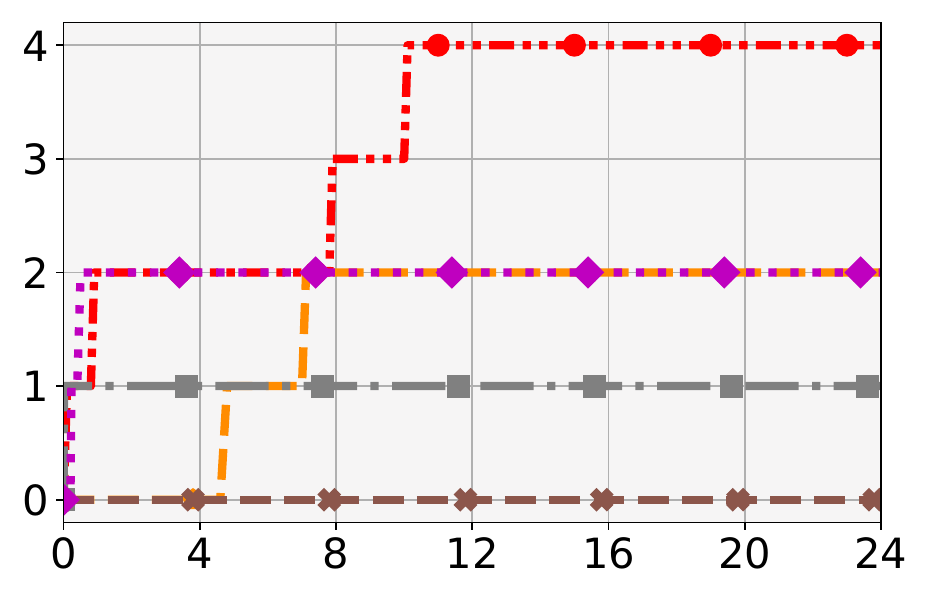}
  \caption{\tidb: detected bugs}
  \label{fig:tidb-bug}
\end{subfigure}
\begin{subfigure}[c]{0.24\textwidth}
  \includegraphics[width=\textwidth]{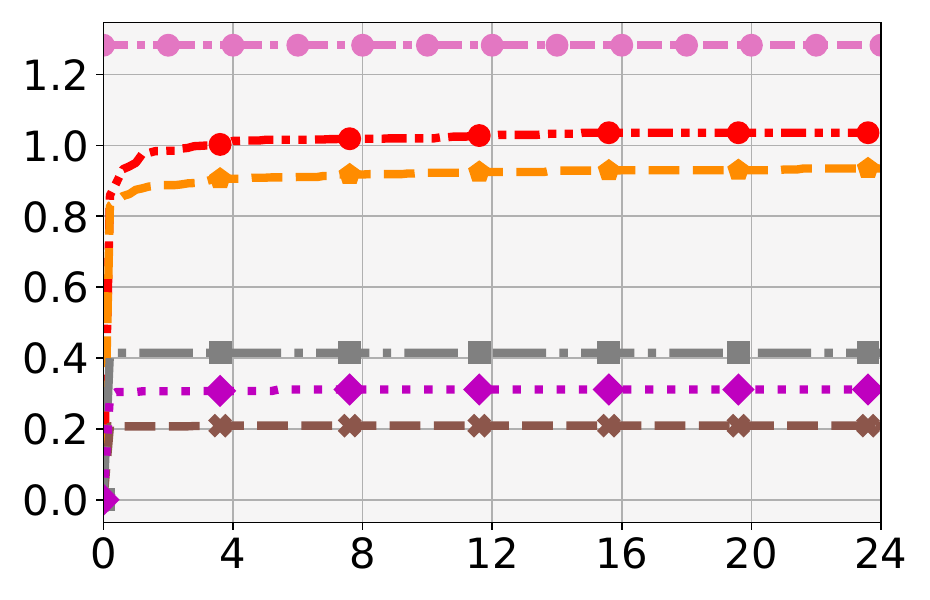}
  \caption{\tidb: grammar edges (k)}
  \label{fig:tidb-gram-size}
\end{subfigure}
\begin{subfigure}[c]{0.24\textwidth}
  \includegraphics[width=\textwidth]{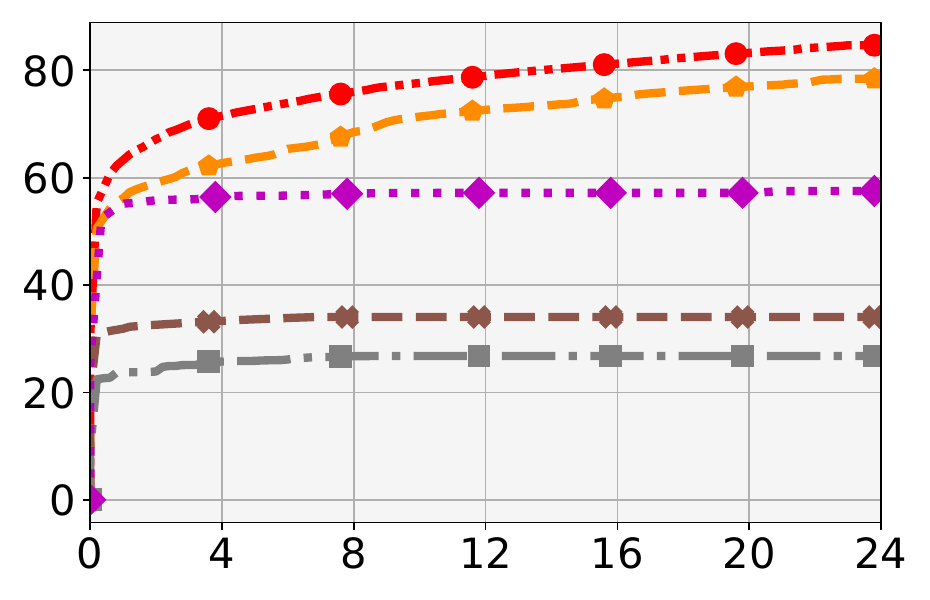}
  \caption{\tidb: code coverage (k)}
  \label{fig:tidb-code-coverage}
\end{subfigure}
\begin{subfigure}[c]{0.24\textwidth}
  \includegraphics[width=\textwidth]{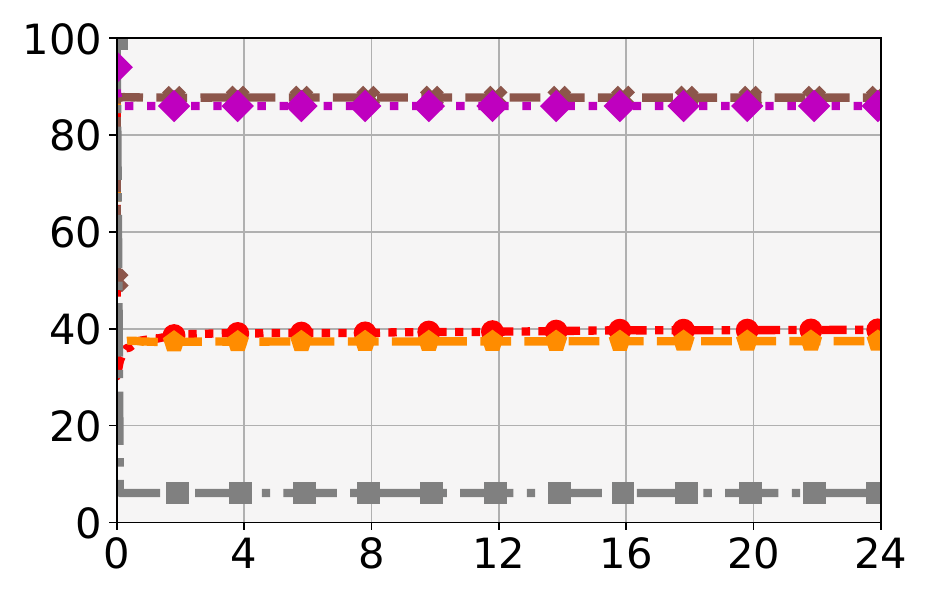}
  \caption{\tidb: query validity (\%)}
  \label{fig:tidb-correct-rate}
\end{subfigure}

\vspace{2.5mm}


\begin{subfigure}[c]{0.24\textwidth}
  \includegraphics[width=\textwidth]{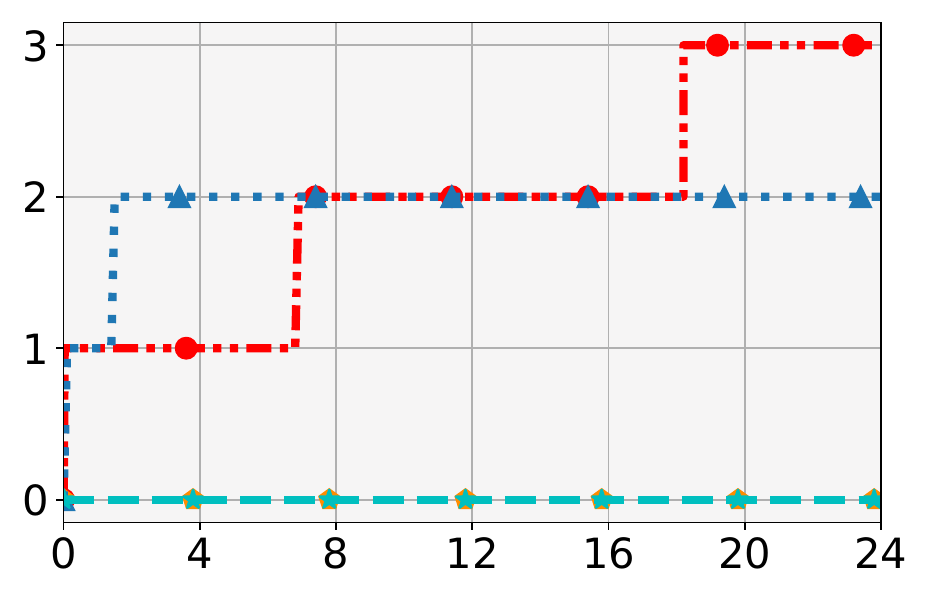}
  \caption{\mariadb: detected bugs}
  \label{fig:mariadb-bug}
\end{subfigure}
\begin{subfigure}[c]{0.24\textwidth}
  \includegraphics[width=\textwidth]{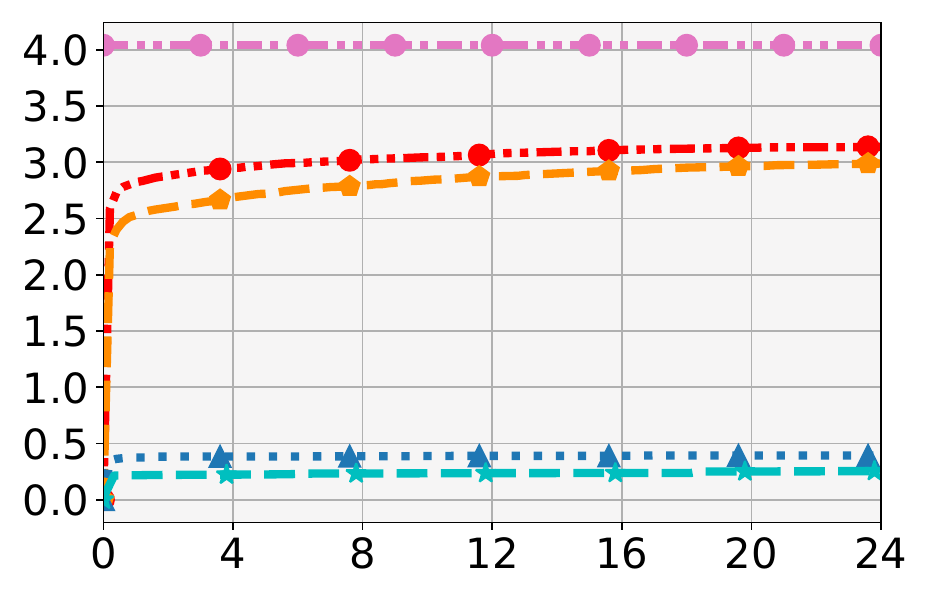}
  \caption{\mariadb: grammar edges (k)}
  \label{fig:mariadb-gram-size}
\end{subfigure}
\begin{subfigure}[c]{0.24\textwidth}
  \includegraphics[width=\textwidth]{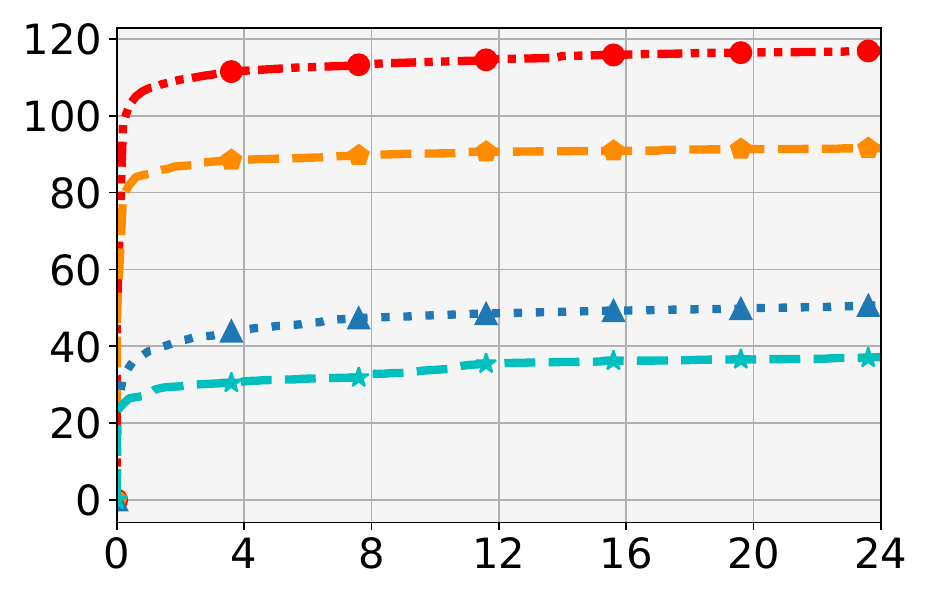}
  \caption{\mariadb: code coverage (k)}
  \label{fig:mariadb-code-coverage}
\end{subfigure}
\begin{subfigure}[c]{0.24\textwidth}
  \includegraphics[width=\textwidth]{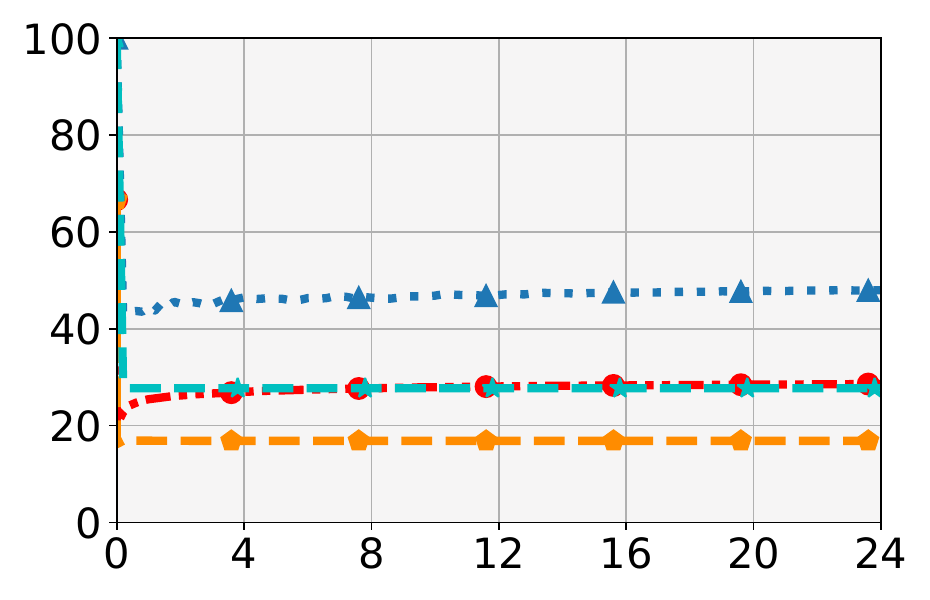}
  \caption{\mariadb: query validity (\%)}
  \label{fig:mariadb-correct-rate}
\end{subfigure}

\vspace{2.5mm}


\begin{subfigure}[c]{0.24\textwidth}
  \includegraphics[width=\textwidth]{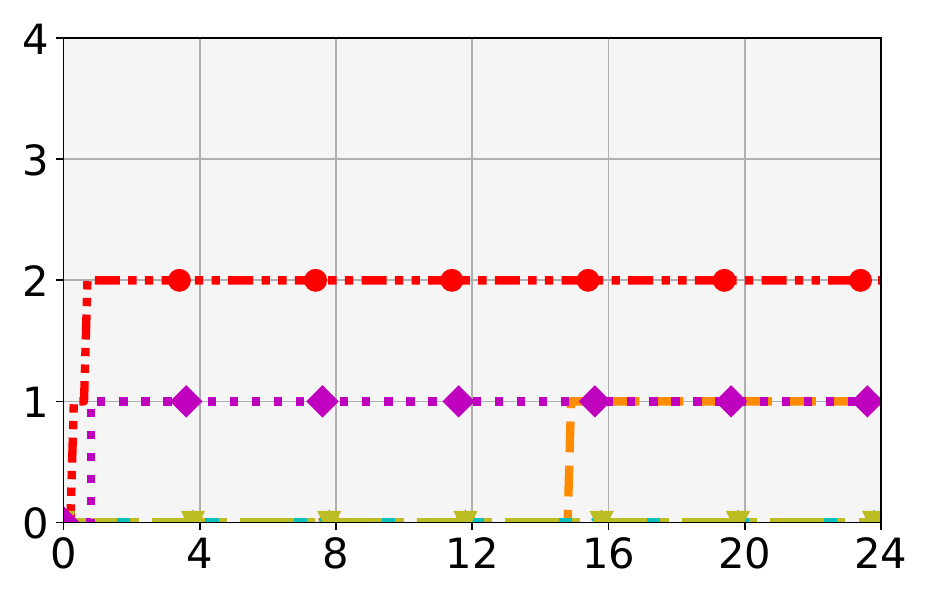}
  \caption{\sqlite: detected bugs}
  \label{fig:sqlite-bug}
\end{subfigure}
\begin{subfigure}[c]{0.24\textwidth}
  \includegraphics[width=\textwidth]{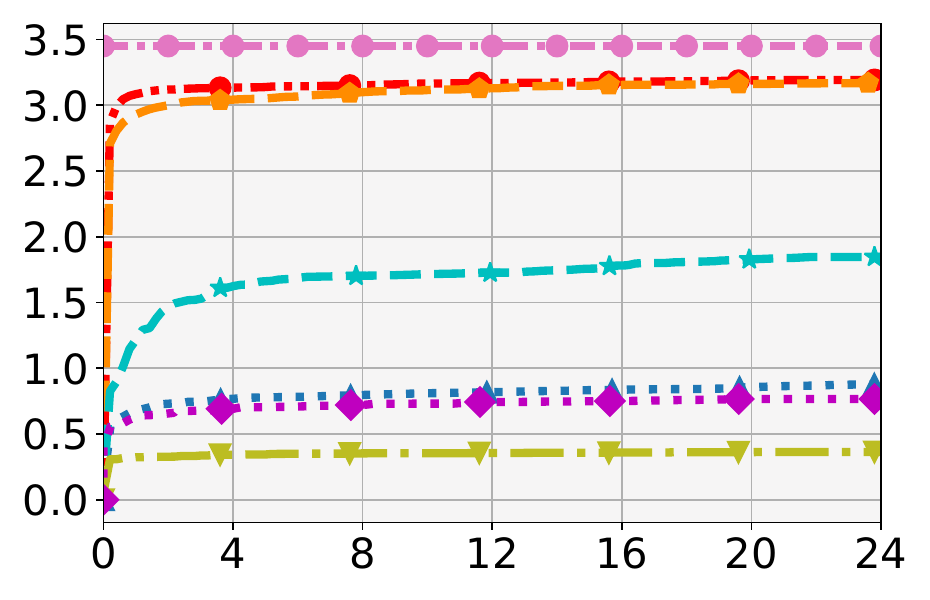}
  \caption{\sqlite: grammar edges (k)}
  \label{fig:sqlite-gram-size}
\end{subfigure}
\begin{subfigure}[c]{0.24\textwidth}
  \includegraphics[width=\textwidth]{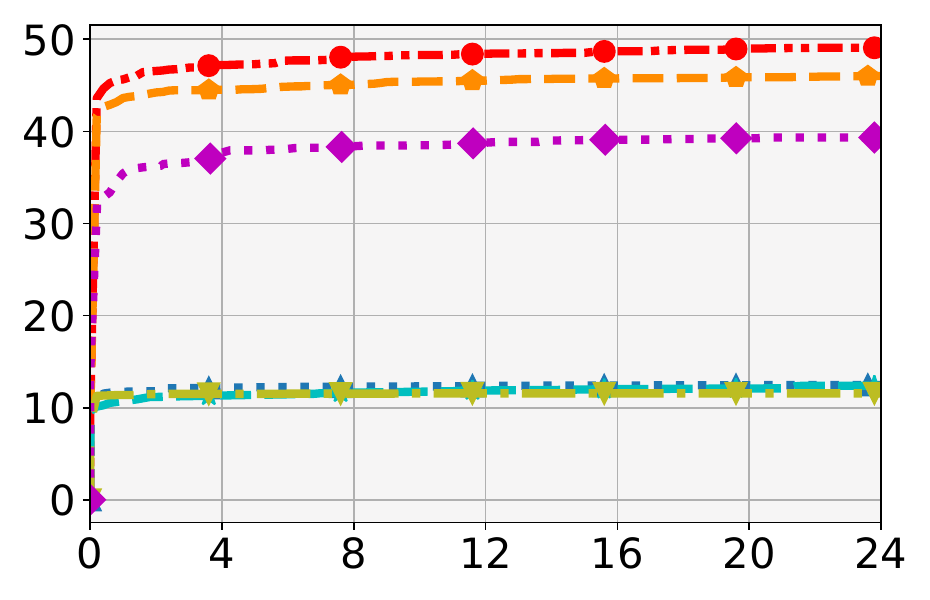}
  \caption{\sqlite: code coverage (k)}
  \label{fig:sqlite-code-coverage}
\end{subfigure}
\begin{subfigure}[c]{0.24\textwidth}
  \includegraphics[width=\textwidth]{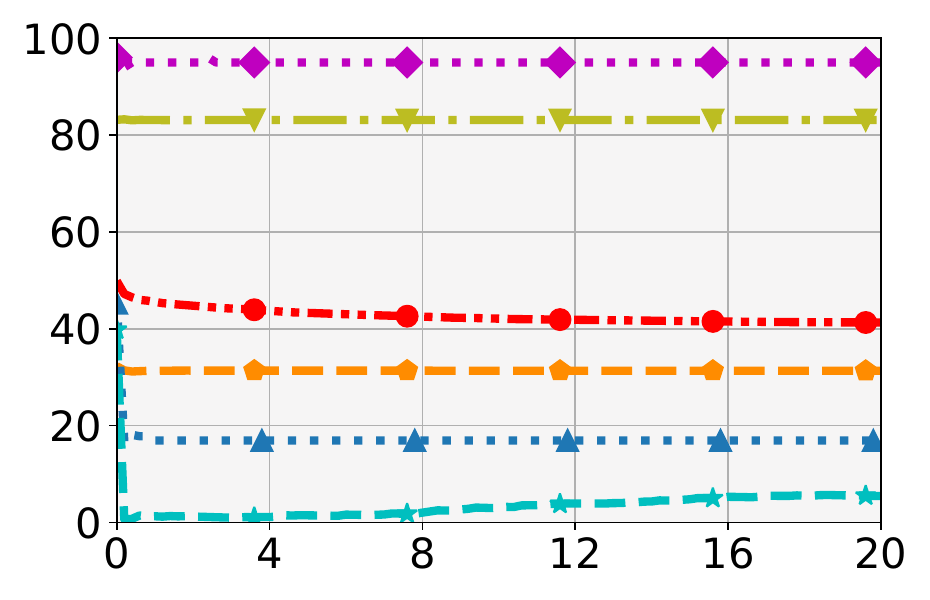}
  \caption{\sqlite: query validity (\%)}
  \label{fig:sqlite-correct-rate}
\end{subfigure}

\caption{Evaluation of different testing tools on \mysql, \cockroachdb, \tidb, \mariadb and \sqlite.}
\label{fig:eval-all-dbms}

\end{minipage}
\end{figure*}

We compare \sys with state-of-the-arts on all 5 supported DBMSs, including \mysql, \sqlite,
\mariadb, \cockroachdb and \tidb. 
For all experiments, we allocate 5 concurrent processes for each tool to stress-test the DBMSs. 
Each evaluation lasts for 24 hours, and we repeat all the experiments 
3 times.
\autoref{fig:eval-all-dbms}
show the results 
we collected.

\PP{Unique bug numbers.}
Across all evaluations conducted on the 5 DBMSs, 
\sys detects the highest number of bugs within 24 hours. 
As seen in~\autoref{fig:mysql-bug}, \sys detects 4 bugs in total, winning the first
place of the evaluation.
\squirrel can also find one new crashing bug from \mysql. 
For the new bug detected from \squirrel, we also reported it to the \mysql developer.
In addition, \squirrel identifies 2 crashing bugs in~\autoref{fig:mariadb-bug}. 
However, the detected bugs from \squirrel are old bugs that had already been known to the 
developer back in 2019 and 2022 respectively. 
%
%
Despite this, \sys records the highest bug count with 3 bugs detected in \mariadb
fuzzing.
Moreover, as shown in~\autoref{fig:cockroachdb-bug} and~\autoref{fig:tidb-bug},
\sys detects remarkable numbers of bugs when testing on \cockroachdb and
\tidb, giving 6 and 4 bugs respectively.
Although \sqlancerqpg can detect multiple logic bugs in \tidb in~\autoref{fig:tidb-bug}, it detects less
memory errors than \sys in all \sqlite, \cockroachdb and \tidb testings.
All baselines tools except \sqlancerqpg do not detect any issues in \sqlite evaluation as shown 
in~\autoref{fig:sqlite-bug}, where \sys detects
2 bugs within the set time frame.

\PP{Grammar edge number.}
The extensive amount of grammar edge triggered by the \sys fuzzing is
the primary reason why it can find more memory errors compared to other baseline tools.
A grammar edge represents the possible combinations between two non-terminal keywords. 
For example, in \autoref{l:ruleprioritization}, a keyword mapping from \cc{table_reference}
to \cc{table_factor} represents one edge case, and \cc{table_reference} to \cc{joined_table} represents
another. 
The upper bound lines display the total possible grammar edges for each DBMSs' grammar rules.
The grammar edge coverage plots are presented in~\autoref{fig:mysql-gram-size},~\autoref{fig:cockroachdb-gram-size},
~\autoref{fig:tidb-gram-size},~\autoref{fig:mariadb-gram-size} and \autoref{fig:sqlite-gram-size}.
While we claim that \sys can account for all the grammar edges from the defined grammar rules, 
the gaps between \sys's grammar edges and the upper bounds indicate the grammar syntaxes we intensionally
exclude.
These syntax features are omitted primarily because they could corrupt the database source,
forcing the DBMS to reboot or disrupting the DBMS server-client connection during the fuzzing loop. 
The excluded grammars include user modification statements, privilege modification statements,
and data read-write lock modifications, among others.
DBMS testers can decide whether to include these syntax elements in their testing. 
But including them would likely reduce the DBMS fuzzing speed.
While \sys captures all the interesting grammar edges in our evaluation, other tools barely match its performance.
The extra syntax elements learned by \sys enables more diverse query generation, resulting in more memory errors reported
than the baseline tools.

\PP{Code coverage.}
\sys reaches the highest DBMS code coverage across all 5 DBMSs' experiments. 
The code coverage plots are shown in~\autoref{fig:mysql-code-coverage},~\autoref{fig:cockroachdb-code-coverage},
~\autoref{fig:tidb-code-coverage},~\autoref{fig:mariadb-code-coverage} and \autoref{fig:sqlite-code-coverage}.
Notably, \sys doesn't rely on any input corpus to reach this level of code coverage, 
sparing the efforts from the DBMS testers to gather interesting queries
as input seeds.

\PP{Query correctness rate.}
The query correctness rate is illustrated in~\autoref{fig:mysql-correct-rate},~\autoref{fig:cockroachdb-correct-rate},
~\autoref{fig:tidb-correct-rate},~\autoref{fig:mariadb-correct-rate} and~\autoref{fig:sqlite-correct-rate}. 
\sys, along with other mutation-based fuzzing tools, generally has a lower query correctness rate 
compared to generation-based tools that rely on hand-written templates.
For example, \sqlancerqpg, \gosqlsmith and \sqlsmith all achieve 
high query validity in their own tests, with \cockroachdbsqlsmith
being an exception.
However, these generation-based tools lack the flexibility to produce diverse query statements, 
because all the generated queries patterns must be hand-written by the developers, making the process 
labor-intensive.
Therefore, \sys can find the highest number of bugs by generating more diverse queries and saturating all the grammar rules
defined for the parsers.

\boxbeg
Overall, \sys can find more memory errors than other testing tools,
because it thoroughly examines all the grammar rules defined 
in the parser, and can reach the highest number of 
grammar coverage upon testing.
Although most generation-based DBMS testers can guarantee a high query 
correctness rate, the requirement for hand-written SQL templates limits
their scalability, resulting in less flexible solutions in the end.
\boxend

\vspace{-4mm}
\subsection{Contribution of Coverage Feedback}
\label{ss:eval-feedback}

To understand the contribution of code coverage, 
we introduce an alternative configuration of \sys, labeled as \sysnocov, to evaluate 
the fuzzer performance without code coverage feedback.
\sysnocov discards all code coverage information obtained from the
DBMS execution, transforming \sys into a pure random 
query generation tool. 
We compare \sys with \sysnocov in all the 5 supported DBMSs. 
The results are embedded in the same plots we used to evaluate on 
different tools in 
\autoref{fig:eval-all-dbms}.
%

\PP{Unique bug numbers.}
Without code coverage guidance, the pure generation-based testing tool \sysnocov
degrades significantly in bug finding capability.
It detects 6 bugs, as opposed to 19 bugs detected by \sys across all DBMSs. 
The largest difference occurs in the \cockroachdb evaluation in~\autoref{fig:cockroachdb-bug}.
\sys detects 6 bugs in total, 
but without code coverage guidance, \sysnocov finds none.

\PP{Grammar edge coverage.}
Interestingly, even if \sys and \sysnocov both share the same grammar definition file, 
\sys achieves a slightly higher grammar edge coverage compared to \sysnocov.
The code coverage feedback aids \sys in exploring hard-to-triggered query syntaxes, 
thereby uncovering more interesting syntax elements which are absent in the
pure query generation. 
%

\PP{Code coverage.}
%
\sys achieves higher code coverage than \sysnocov
as anticipated. 
Simply covering all syntax elements from the grammar definition file doesn't provide a complete picture.
But smartly combines different elements together to 
form different interesting contexts is also crucial for generating more interesting test cases.
Code coverage feedback guides the fuzzer to gradually accumulate interesting syntax 
elements together in the fuzzing queue.
The various combinations between different DBMS
features lead to more unexpected SQL contexts for the DBMS handling logic, 
and eventually lead to more bugs detected.

\boxbeg
The code coverage feedback accumulates the interesting syntax features discovered from the grammar 
rule definition file, and it creates more interesting feature combination contexts.
It helps \sys to achieve higher code coverage, 
and eventually leads to more bugs detected.
\boxend

\subsection {Contribution of Diverse Syntax Features}
\label{ss:eval-syntax}

To demonstrate how vast query syntaxes enhance bug-finding, 
we refer to~\autoref{t:all-segfaults} and~\autoref{t:all-asserts} to show the benefits.
Assuming infinite resources can be allocated,
the columns of \squirrel, \sqlsmith, \cockroachdbsqlsmith, \gosqlsmith and \sqlancerqpg 
in~\autoref{t:all-segfaults} and~\autoref{t:all-asserts}
indicate whether each bug could theoretically be detected by referenced tools. 
Given the input corpus and Internal Representation (IR),
\squirrel can only detect 
8 out of 37 bugs that \sys found. 
\sqlsmith can detect half of the bugs reported by \sys (3 out of 6).
\cockroachdbsqlsmith can detects 13 out of 37 bugs from \cockroachdb. 
\gosqlsmith can detect none of the bugs from \sys.
\sqlancerqpg can detect 11 out of 50. 
The diverse syntaxes enable \sys to explore more interesting features from the DBMSs, 
and trigger more interesting bugs that are overlooked by these baseline tools.
Next,
we present two case studies to demonstrate the uniqueness of bugs detected by \sys.

\begin{figure}[t]
\begin{lstlisting}[style=mystyle,label={l:case1},caption={\textbf{A one-line query that crashes \tidb}, which aims to recover a table from a non-existing DDL JOB ID.}]
RECOVER TABLE BY JOB 0;
\end{lstlisting}
\vspace{-2mm}
\end{figure}

\PP{One-line query that crashes TiDB}
\autoref{l:case1} presents a unique bug from \tidb. 
The PoC is surprisingly simple, consisting of just one line
of SQL query. 
But the simple PoC crashes the \tidb query executor, and results in
an immediate loss of connection between the \tidb server and client. 
The PoC attempts to recover a table that had been previously dropped
from the database.
%
A more commonly use case is to directly recover the table by its table name, \ie, 
using \cc{RECOVER TABLE table_0;} to bring back the deleted table 
\cc{table_0}. 
However, in conner cases where the DBMS user has created another table that share
the same name as the deleted one, \tidb offers an alternative form of the \cc{RECOVER} statement,
as shown by \autoref{l:case1}, 
that uses \cc{DDL JOB ID} to recover the table that were previously removed.
The \cc{DDL JOB ID} information can be fetched by using the \cc{ADMIN SHOW DDL JOBS;} statement,
where the \cc{DDL JOB ID} saves the unique ID for all Data Definition Language (DDL)
after their executions.
Interestingly, the \cc{DDL JOB ID} value can never be 0. 
But the parser from \tidb never checks the value from the PoC, 
and the value \cc{DDL JOB ID} equals 0 is successfully set in the \tidb backend.
%
In this case, 
\tidb interprets the statement as \cc{RECOVER TABLE table_name} and then call the
\cc{getRecoverTableByTableName} function.
Unfortunately, the table name variable is remain uninitialized, resulting in a nil pointer dereference bug
from \tidb and crashes the \tidb worker process.
This bug is interesting and was never detected before because the \cc{RECOVER TABLE BY JOB} 
statement has been rarely tested.
Since the feature introduced after \tidb version 3.0, there are only 13 instances in the 
\tidb unit tests reference this feature.
Furthermore, all the unit tests are constructed with pre-defined or fixed \cc{DDL JOB ID}, which are not helpful
to trigger this bug.
Additionally, the official query generation-based testing tool from the \tidb developer, \gosqlsmith, 
does not support this grammar feature in its query generation templates.
Our tool \sys directly parses the grammar rule definition file designed for \tidb, and automatically
recognizes the \cc{RECOVER TABLE BY JOB} syntax grammar. 
\sys then generates the \cc{RECOVER TABLE BY JOB} statement, and fills in the \cc{DDL JOB ID}
with arbitrary integer, such as value 0, and eventually triggers this bug.

\begin{figure}[t]
\begin{lstlisting}[style=mystyle,label={l:case2},caption={\textbf{A \mariadb crash that corrupts the database source.}}]
CREATE TABLE IF NOT EXISTS t0 (c1 INT) PARTITION BY 
  HASH(c1);
ALTER TABLE t0 CHECK PARTITION ALL FOR UPGRADE;
ALTER TABLE t0 ORDER BY c1;
\end{lstlisting}
\vspace{-2mm}
\end{figure}

\PP{A database corruption bug from MariaDB.}
\autoref{l:case2} shows a segmentation fault PoC from \mariadb DBMS.
The \cc{CREATE TABLE} statement creates a table \cc{t0} with one column
\cc{c1}.
The table is partitioned by \cc{HASH(c1)}.
Table partitioning is used to split one table data into multiple subsets, and store them
individually to ease their managements or speed up their access speed. 
The \cc{PARTITION BY HASH} directive tells the DBMS to handle the data partitioning, 
and make sure the data are distributed evenly in the split partitions.
The second statement runs \cc{ALTER TABLE CHECK PARITION} on the just created table.
It verifies whether the created partitions contain any errors. 
The additional syntax \cc{FOR UPGRADE} checks if the current partitions are compatible with the currently
running \mariadb version.
Right after the partition checking, the third statement modifies the table \cc{t0} and 
reorders the table contents based on the data in column \cc{c1}.
By running all three statements together, \mariadb crashes with corrupted memory access. 
What's worse, the PoC also corrupts the \cc{DDL_LOG} section from the database source,
causing future \mariadb crashes whenever \mariadb accesses this database. 
We have reported the PoC to the \mariadb developers and they are working on the patch.

This bug can only be detected by \sys in our evaluation.
The crucial step in triggering this bug is to combine \cc{CHECK PARTITION ALL} with \cc{FOR UPGRADE},
and then call \cc{ALTER TABLE} immediate after. 
However, the syntax features of \cc{CHECK PARTITION ALL} and \cc{FOR UPGRADE} are rarely touched by the 
existing testing tools.
\squirrel doesn't support either syntaxes in its internal
parser. 
These two syntax features are absent in the \squirrel's input corpus either.
Furthermore, the combined usage of \cc{CHECK PARTITION ALL} and \cc{FOR UPGRADE} is also 
not presented in the \mariadb official unit test library,
where \cc{FOR UPGRADE} is more commonly used for \cc{CHECK TABLE} in the test 
instead of \cc{CHECK PARTITION}. 
Our tool \sys does not rely on any input corpus to realize the different syntax features,
where the input corpus are often gathered from the DBMS's unit test library.
Instead, \sys learns the syntax rules from \mariadb's built-in parser, and automatically
constructs queries that contain the \cc{CHECK PARTITION ALL} and \cc{FOR UPGRADE} symbols. 
Therefore, \sys is the only tool we tested that can detect this bug from \mariadb.

\boxbeg
The diverse syntax features learned from grammar definition files enable \sys
to generate more diverse testing queries, which cover more interesting syntax features from the DBMSs, and 
therefore brings more interesting bugs that are not possible from the previous tools.
\boxend

\section{Discussion}
\label{s:discussion}

\begin{figure}[t]
\begin{lstlisting}[style=onlycommenthl, label={l:syntaxruleinparser},caption={\textbf{BNF grammar does not tell the whole story.} 
  We simplify these rules from \sqlite's parser and covert them to BNF.}]
joinop: JOIN_KW JOIN_SYM
      | JOIN_KW nm JOIN_SYM
      | JOIN_KW nm nm JOIN_SYM ;
nm:     IDENTIFIER /* terminal keyword for query arg */
      | JOIN_KW;   /* terminal keyword from tokens */
/* Mapped query tokens for JOIN_KW
 * "CROSS"   -> "JOIN_KW"
 * "FULL"    -> "JOIN_KW"
 * "INNER"   -> "JOIN_KW"
 * "LEFT"    -> "JOIN_KW"
 * "NATURAL" -> "JOIN_KW"
 * "OUTER"   -> "JOIN_KW"
 * "RIGHT"   -> "JOIN_KW"
 */
\end{lstlisting}
\end{figure}

\PP{Syntax rules outside grammar definition file.}
\sys delivers a promising result in exploring syntax elements
defined in SQL grammar definition files.
However,
some grammar rules can be pushed down to the DBMS
backend, not in the grammar definition file.
For example,
\sqlite contains a parser generator tool, \cc{Lemon}~\cite{sqlite-parser},
which supports grammar definition syntax similar to BNF
that \cc{Yacc} and \cc{Bison} support.
However, the grammar definition file provided to \cc{Lemon}
does not strictly represent the ground truth grammar 
complied by the \sqlite frontend,
since many grammar checks are pushed down to the \sqlite back-end to handle.
\autoref{l:syntaxruleinparser} shows one case that the parser rule in `parser.y' cannot faithfully defines all the syntax
constraints from the \sqlite's parser.
The example focuses on the \cc{joinop} keyword, which is used to combine data from two or more tables. 
The valid query segments matching \cc{joinop} could be `LEFT JOIN', `RIGHT JOIN', and `LEFT INNER JOIN', `RIGHT OUTER JOIN' etc.
As we can see the terminal keyword \cc{JOIN_KW} can map to all the tokens we mention here, so all these valid cases should
pass the grammar check. 
However, the three rules defined in \cc{joinop} do not enforce the order of the \cc{JOIN_KW} tokens, which means 
a query such as `LEFT RIGHT JOIN' might also pass the grammar check, and later turns out to be invalid. 
What's worse, the \cc{nm} keyword brings in \cc{IDENTIFIER} as an alternative choice that we can fill into the \cc{joinop} rules,
which would also be rejected by \sqlite parsing in the end.
It turns out \sqlite implements a function named \cc{sqlite3JoinType}, that are designed to verify the keyword contents passed 
into the \cc{joinop} grammar rules.
It effectively rejects any \cc{IDENTIFIER} keywords in the parsed syntax tree.
It also rejects corner cases such as `LEFT RIGHT JOIN'. 
However, these additional logics are written in \cc{C} language instead of using BNF grammar define notation. 
So there is no common pattern we can make used of to acknowledge these extra syntax restrictions. 
Therefore, we currently rely on human efforts to mark these hidden grammar constraints. 
For instance, we replace all the \cc{nm} keywords to \cc{JOIN_KW} in \cc{joinop}, 
and filter out any `LEFT RIGHT JOIN' in the query generation.

\section{Conclusion}
\label{s:conclusion}

We design \sys,
a novel fuzzing tool that
automatically extracts syntax features
from DBMSs built-in grammar definition files.
By traversing these grammar files,
\sys explores all grammar rules defined in each DBMS, 
and thus can generate more diverse testing queries
than previous DBMS testing tools.
\sys detects \totalbugs bugs
across five popular DBMSs,
\sqlite, \mysql, \cockroachdb, \tidb and \mariadb.
The evaluation shows
\sys achieves the highest grammar coverage,
the highest code coverage and
reports more bugs within 24-hour experiments.
We have reported all detected bugs to the corresponding DBMS developers. 
They have confirmed all the bugs and fixed \fixed of them.

\balance
{
\bibliographystyle{plain}
\footnotesize
\bibliography{main}
}

\end{document}